\documentclass[preprint,3p,twocolumn]{elsarticle}
\usepackage{geometry}
\usepackage{fancyhdr}
\biboptions{numbers,sort&compress}

\usepackage{amssymb}
\usepackage{amsmath}
\usepackage{url}

\biboptions{comma,square,sort&compress}

\begin{document}

\begin{frontmatter}


\title{Mapping of Microstructure Transitions during Rapid Alloy Solidification using Bayesian-Guided Phase-Field Simulations}

\author[a,b]{Jos\'e Mancias}
\author[a]{Brent Vela}
\author[c]{Juan Flórez-Coronel}
\author[b]{Rouhollah Tavakoli}
\author[c]{Douglas Allaire}
\author[a,c,d]{Raymundo Arr\'oyave}
\author[b]{Damien Tourret}

\affiliation[a]{organization={Department of Materials Science and Engineering},
            addressline={Texas A\&M University}, 
            city={College Station},
            postcode={77843}, 
            state={TX},
            country={USA}}

\affiliation[b]{organization={IMDEA Materials},
            city={Getafe},
            postcode={28906}, 
            state={Madrid},
            country={Spain}}

\affiliation[c]{organization={J. Mike Walker ’66 Department of Mechanical Engineering},
            addressline={Texas A\&M University}, 
            city={College Station},
            postcode={77843}, 
            state={TX},
            country={USA}}

\affiliation[d]{organization={Wm Michael Barnes '64 Department of Industrial and Systems Engineering},
            addressline={Texas A\&M University}, 
            city={College Station},
            postcode={77843}, 
            state={TX},
            country={USA}}

\begin{abstract}

This study addresses microstructure selection mechanisms in rapid solidification, specifically targeting the transition from cellular/dendritic to planar interface morphologies under conditions relevant to additive manufacturing. 
We use a phase-field model that quantitatively captures solute trapping, kinetic undercooling, and morphological instabilities across a broad range of growth velocities ($V$) and thermal gradients ($G$), and apply it to a binary Fe-Cr alloy, as a surrogate for 316L stainless steel.
By combining high-fidelity phase-field simulations with a Gaussian Process-based Bayesian active learning approach, we efficiently map the microstructure transitions in the multi-dimensional space of composition, growth velocity, and temperature gradient. 
We compare our PF results to classical theories for rapid solidification.
The classical KGT model yields an accurate prediction of the value of $G$ above which the interface is planar for any growth velocity.
Microstructures transition from dendrites to cells as the temperature gradient increases close to this value of $G$.
We also identify the occurrence of unstable ``intermediate'' microstructures at the border between dendritic and planar at low $G$, in the absence of banding instability in this Fe-Cr alloy.
Our results highlight the capabilities of Bayesian-guided PF approaches in exploring complex microstructural transitions in multidimensional parameters spaces, thereby providing a robust computational tool for designing process parameters to achieve targeted microstructures and properties in rapidly solidified metallic alloys.

\end{abstract}


\end{frontmatter}

\thispagestyle{fancy} 

\section{Introduction} \label{intro}

The emergence of additive manufacturing (AM) has sparked a renewed interest in rapid solidification and resulting microstructures, which often strongly affect resulting material properties and performance \cite{konig2023solidification, wang2016effect, shi2020microstructural, liu2022additive}.
Theoretical foundations for rapid solidification are relatively well established, and underlying mechanisms, such as solute trapping, kinetic undercooling, and the morphological stability of the solid-liquid interface, are well identified (see Section~\ref{sec:rapid} \cite{dantzig2016solidification,kurz2023fundamentals}).
However, the dynamical phenomena of morphological transitions between microstructural patterns at high growth velocity, e.g., close to the transition between dendritic/cellular and planar interface stability, remain relatively poorly explored.

\begin{figure}[b!]
  \centering
  \includegraphics[width=\columnwidth]{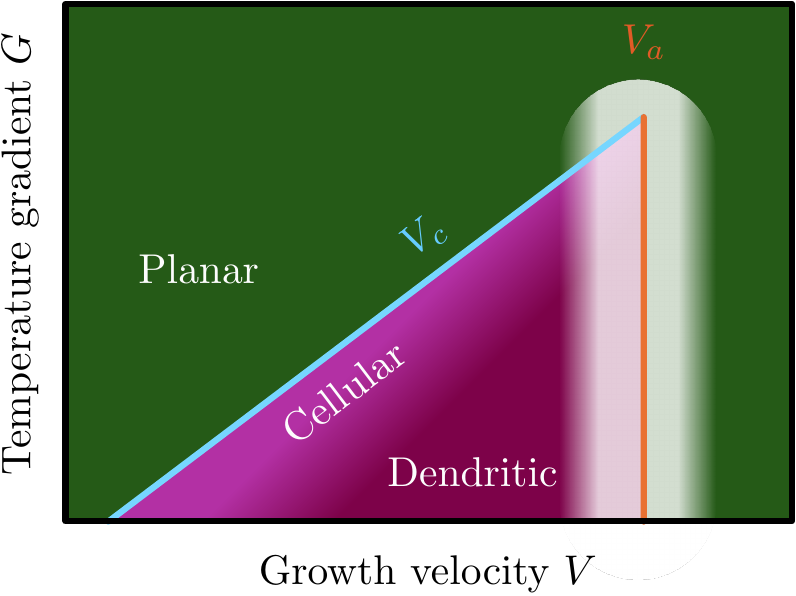}
  \caption{
  Typical microstructure selection map in the $(G,V)$ plane.
  The region of interest, in the vicinity of the absolute stability limit $V_a$, is the faded ($\sim$ white) region. 
  }\label{fig:gv_schem}
\end{figure}

The overall microstructure selection map for an alloy is typically illustrated schematically, as shown in Figure~\ref{fig:gv_schem}. 
At low-to-moderate solid-liquid interface velocity, $V$, the transition from a planar front to a more morphologically complex interface pattern occurs at a velocity $V_c$, prescribed by the constitutional undercooling criterion \cite{tiller_redistribution_1953} or, more accurately, the Mullins-Sekerka stability analysis \cite{mullins_stability_1964}.
From an experimental standpoint, the transition from planar to cellular to dendritic growth regimes at $V\approx V_c$ and above, is well understood and documented, based on more than half a century of {\it in situ} observation of organic compounds solidification experiments \cite{jackson1965transparent, losert1998evolution, akamatsu2016situ}, and decades of {\it in situ} observations of metallic alloys using X-ray imaging \cite{mathiesen1999time, yasuda2004direct, nguyen2012interest, clarke2017microstructure}.

At very high $V$, planar growth of the solid-liquid interface is also known to be stable.
The absolute stability threshold, $V_a$, is often considered independent of the temperature gradient $G$ \cite{mullins_stability_1964, trivedi1989solidification}.
It is also known that this threshold may be better estimated by the velocity, here denoted $V_{\max\{T_S\}}$, at which the alloy solidus temperature, $T_S(V)$, exhibits a maximum due to competition between solute trapping and kinetic undercooling effects \cite{tourret2023morphological, karma1993interface, kurz1996banded}.
This was supported by recent laser spot melting experiments of a ternary Ni alloy with {\it in situ} synchrotron X-ray imaging enabling the accurate measurement of growth velocities, with dendritic patterns identified in regions grown at $V_a<V<V_{\max\{T_S\}}$.

Schematic representations like Fig.~\ref{fig:gv_schem} sometimes feature a cellular region along the $V_a$ limit, by analogy to the transition observed at $V\approx V_c$.
However, the presence of a cellular growth regime therein is -- to the best of our knowledge -- not clearly established.
Instead, it appears that ``cellular'' patterns observed in rapidly solidified (e.g., AM) micrographs might in fact be dendritic patterns devoid of sidebranches \cite{tourret2023morphological}.
For some alloys, in particular Al alloys \cite{carrard_about_1992, gill1993laser, kurz1996banded, mckeown2016time}, and also Mg alloys \cite{tourret_emergence_2024}, this region is known to produce an oscillatory instability leading to ``banded'' microstructures.
Other prominent alloy systems, like Fe- or Ni-based alloys, were -- to the best of our knowledge -- never reported to exhibit such banded microstructures. 
Their transition from dendritic to planar morphology, e.g., regarding the presence or absence of a cellular regime, is not clearly identified.

In this context, our main objective is to provide a deeper understanding of microstructure selection in this intermediate region between dendritic and planar regimes at high growth rate (i.e., the washed out region in Fig.~\ref{fig:gv_schem}) in alloys that do not exhibit an oscillatory banding instability.
Specifically, we focus on a Fe-Cr alloy, used as a surrogate for a 316L stainless steel, and focus on a $(G,V)$ parameter range relevant to fusion-based AM processes.

The relative lack of knowledge on microstructure selection at high growth rate is, to a great extent, due to (i) the difficulty to observe interfaces {\it in situ} at a relevant high $V$, but also (ii) the lack of computational models capable of operating quantitatively throughout the relevant velocity range at relevant length and time scales.
In the context of solidification, ``quantitative'' phase-field (PF) models refer to models that remain faithful to the sharp-interface free-boundary problem, typically defined by the combination of solute and/or heat transport (e.g. diffusion) in bulk phases, the Stefan condition for heat/solute conservation at the interface, and the Gibbs-Thomson condition describing the equilibrium, or departure therefrom, at the interface.
In fact, classical PF models for solidification are capable of quantitative capture of rapid solidification mechanisms, e.g., solute trapping \cite{ahmad_solute_1998, danilov_phase-field_2006, galenko_solute_2011}, but to do so they require the use of a diffuse interface width $W$ close to the actual width of the solid-liquid interface $W_0$.
This limitation to $W\approx W_0$ severely restricts the affordable numerical spatial discretization, and hence the application of such models to the relevant microstructural length scale.
Upscaled quantitative PF models, designed to remain accurate when $W> W_0$, based on asymptotic thin interface analyses \cite{karma1998quantitative, echebarria2004quantitative}, have been widely used in the low-to-moderate regime $V$. 
However, until recently, similar upscaled models were missing for high growth rates.

Among the state-of-the-art rapid solidification PF models, the so-called finite interface dissipation model \cite{steinbach2012phase} provides a great option to explore the rapid solidification regime at the microstructural scale \cite{karayagiz2020finite}.
However, its predictions heavily rely on an elusive interface permeability parameter that is difficult to quantify and therefore is often used as a calibration parameter \cite{karayagiz2020finite,reuther2019solute}.
Other recent approaches \cite{pinomaa2019quantitative, kavousi2021quantitative} are based on a phenomenological modification of the classical anti-trapping current \cite{karma2001phase,echebarria2004quantitative}.
They are calibrated to match, at first order, the classical continuous growth model (CGM) \cite{aziz1982model, aziz1988continuous, boettinger1986science} at the early onset of solute trapping, and therefore, their quantitative nature across the entire relevant velocity range -- from equilibrium partitioning up to complete solute trapping -- is not guaranteed.
Addressing these limitations, Ji et al. \cite{ji_microstructural_2023,ji_quantitative_2024} recently proposed an original PF model, which was shown to quantitatively capture the effects of solute trapping and kinetic undercooling across the entire $V$ range.
This model was found to accurately reproduce high-$V$ banding instability in Al \cite{ji_microstructural_2023,ji_quantitative_2024} and Mg \cite{tourret_emergence_2024} alloys, thereby supporting its capability to appropriately and quantitatively capture key rapid solidification mechanisms in binary alloys.

Furthermore, even when a quantitative PF model is available, the computational cost of individual simulations remains high.
This has motivated the development of computational acceleration strategies such as parallelization \cite{shimokawabe2011peta, vondrous2014parallel, shibuta2015solidification, sakane2022parallel}, adaptive mesh refinement \cite{provatas1999adaptive, greenwood2018quantitative}, advanced time stepping schemes \cite{rosam2008adaptive, bollada2015three, boccardo2023efficiency}, and custom spectral solvers \cite{boccardo2023efficiency, chen1998applications, feng2006spectral}.
Yet, despite their tremendous potential in upscaling simulations, they still fall short when the parameter space to explore is not only broad but also highly multidimensional.
This may be the case, for instance, for multicomponent alloys, which add one extra dimension in the search space for every solute species. 
In such cases, brute-force mapping, such as full factorial design of experiments, is too inefficient, and more advanced exploration strategies are direly required.
In this context, our second aim is to design a Bayesian-guided exploration strategy based on Gaussian processes (GP) to classify interface patterns in rapid solidification. 
We demonstrate the viability and efficiency of this original approach by applying it to the search of the dendritic-to-planar limit in the $(G,V,c_\infty)$ space, with $c_\infty$ the nominal alloy composition. 

\section{Methods}

\subsection{Reminders of solidification theory} \label{sec:rapid}

In solidification theory, rapid solidification refers to a regime in which the solid-liquid interface substantially departs from equilibrium due to reaching a high growth velocity $V$. 
The relevant velocity range depends on the alloy system and even sometimes on individual solute species in a given alloy. Still, the usual order of magnitude for conventional metallic alloys is $V\approx1$~m/s or higher.

In this regime, the solid phase may grow with a solute concentration beyond its equilibrium solubility limit. This phenomenon is known as solute trapping. As a result, the ratio of the interface solute concentration in the solid $c_s$ and liquid $c_l$, known as the solute partition coefficient $k_V = c_s/c_l$, deviates from its equilibrium value, $k_e$, and approaches $1$ as $V$ approaches a characteristic diffusion velocity $V_d$, typically of the order of 1\:m/s. As this happens, the liquidus and solidus lines of the phase diagram converge toward the so-called T$_0$\textendash line as $V$ increases. The velocity dependence of the interface solute partition coefficient, $k_V$, and of the liquidus slope, $m_V$, here considering $k_V\leq1$ and $m_V>0$, are commonly described using the CGM \cite{aziz1982model, aziz1988continuous, boettinger1986science}
\begin{equation} \label{eq_kV}
k_V = \frac{k_e + V/V_d}{1+V/V_d}
\end{equation}
\begin{equation} \label{eq_mVme}
\frac{m_V}{m_e} = \frac{1 - k_V + [k_V + (1-k_V)\alpha] \ln{[(k_V/k_e)]}}{1-k_e}
\end{equation}
where $k_e$ and $m_e$, respectively, represent the equilibrium values of the partition coefficient $k_V$ and of the liquidus slope $m_V$, relevant when $V \ll V_d$.
The coefficient $\alpha$ is a term that accounts for the resistance to movement of the solid-liquid interface, known as solute drag. 
Here, for the sake of consistency with the chosen phase-field model (presented later) \cite{ji_quantitative_2024}, the solute drag coefficient is considered as $\alpha = 0.645$, and the diffusion velocity 
\begin{equation}
    V_d \approx 0.356 \,\frac{\ln (1/k_e)}{(1-k_e)} V_d^0
\end{equation}
where $V_d^0 = D_L / W_0$ with $D_L$ the liquid solute diffusivity and $W_0$ is the actual physical thickness of the solid-liquid interface. 

The interface velocity also affects its growth temperature through a kinetic undercooling. This phenomenon results in an interface temperature, $T$, decreasing as $V$ increases. The dependence on velocity of the liquidus ($T_L$) and solidus ($T_S$) temperatures are thus described by the following equations in the case of a planar interface (i.e. when curvature undercooling is neglected)
\begin{equation} \label{tliquidus}
T_L(V) = T_M - m_Vc_\infty - \frac{V}{\mu_k}
\end{equation}
\begin{equation} \label{tsolidus}
T_S(V) = T_M - \frac{m_V}{k_V} c_\infty - \frac{V}{\mu_k}
\end{equation}
where $c_\infty$ represents the nominal alloy solute concentration, $\mu_k$ is the interface kinetic coefficient, and $T_M$ is the melting temperature of the solvent. Solute trapping can be investigated by compositional analysis of rapidly solidified samples \cite{kittl2000complete}. However, measuring kinetic undercooling experimentally is much more challenging as it requires establishing a correlation between interface velocity and temperature under conditions of deep undercooling \cite{willnecker1989evidence, lum1996high}. 

A clear indicator of rapid solidification is the transition from a dendritic to a planar interface, occurring when the velocity exceeds the so-called absolute stability threshold, $V_a$ \cite{mullins_stability_1964, trivedi1989solidification}. 
It may be calculated via an extension of the Mullins and Sekerka linear perturbation theory to high undercooling as
\begin{equation} \label{eq:absolutestability}
V_a \approx \frac{D_L m_V c_\infty [1-k_V]}{k_V^2 \Gamma_{\rm sl}}
\end{equation}
where $\Gamma_{\rm sl}$ is the Gibbs-Thomson coefficient of the solid-liquid interface. 
Note that, in Eq.~\eqref{eq:absolutestability}, the velocity-dependent $k_V$ and $m_V$ make Eq.~\eqref{eq:absolutestability} an implicit definition of $V_a$.
This equation only provides an order of magnitude of the actual $V_a$, which is usually better estimated via the maximum value of $T_S(V)$ from Eq.\,\eqref{tsolidus} \cite{carrard1992banded, karma1993interface, kurz1996banded, tourret2023morphological}. 
Microstructures in the regime of $V$ just below this maximum, within the region where $dT_S/dV > 0$, may exhibit oscillatory growth between cellular/dendritic microstructures and planar microstructure leading to the formation of ``bands'' normal to the temperature gradient \cite{carrard1992banded, karma1993interface, kurz1996banded}. 
Such banded microstructures, commonly observed in Al alloys, have, to the best of our knowledge, never been reported in Fe- or Ni-based alloys.

As illustrated in Fig.~\ref{fig:gv_schem}, a planar solid-liquid interface is morphologically stable at a growth velocity $V$ either higher than the absolute stability threshold $V_a$ or lower than the constitutional undercooling velocity $V_c$ \cite{tiller_redistribution_1953, mullins_stability_1964} with
\begin{equation} \label{eq:const_underc}
V_c = \frac{D_L G \,k_V}{m_V(1-k_V)c_\infty} \approx \frac{D_L G \,k_e}{m_e(1-k_e)c_\infty} 
\end{equation}
In either cases, as the interface is planar, solute conservation prescribes that, at steady state, the interface stabilizes within the temperature gradient at the solidus temperature.

In a growth velocity range $V_c<V<V_a$, the interface pattern, cellular at $V\approx V_c$ and dendritic for $V\gg V_c$, grows below $T_L$ at a given undercooling strongly dependent on the growth velocity.
This undercooling, and the corresponding dendrite tip radius $R$, can be estimated by combining a statement of solute diffusion in the liquid, e.g. the Ivantsov solution \cite{gp_ivantsov_notitle_1947} prescribing the product $RV$, and an auxiliary condition, e.g. solvability theory \cite{barbieri1989predictions} or more approximately marginal stability \cite{mullins_stability_1964, langer1977stability, langer1978theory} prescribing the product $R^2V$.
A well-known example of such model is the KGT (Kurz, Giovanola, Trivedi) model \cite{kurz_theory_1986, rappaz_analysis_1990}, which may be written as 
\begin{equation} \label{KGT_main}
\frac{4\pi^2 \Gamma_{\rm sl}}{R^2} + \frac{2}{R} 
\frac{\text{Pe}\, m_V c_0 (1 - k_V) \xi (\text{Pe}, k_V)}
{1 - \left[ (1 - k) {\rm Iv}({\rm Pe}) \right]}
+ G = 0
\end{equation}
with
\begin{equation} \label{KGT_second}
    \xi({\rm Pe}, k_V) = 1-\frac{2k_V}{\sqrt{1+(2\pi/{\rm Pe})^2} - 1 + 2k_V}
\end{equation}
The P\'eclet number is defined as ${\rm Pe} = RV/(2D)$ and the tip temperature is 
\begin{equation} \label{KGTTipTemperature}
T = T_M + \frac{m_V c_\infty}{1-(1-k_V){\rm Iv}({\rm Pe})} - \Gamma_{\rm sl}\kappa
\end{equation}
with $\kappa$ the tip curvature and ${\rm Iv}({\rm Pe})$ the Ivantsov solution of the steady diffusion field in front of a parabolic tip \cite{gp_ivantsov_notitle_1947}.
The KGT model is typically used in three-dimensions (3D), where $\kappa=2/R$ and ${\rm Iv}_{\rm 3D}({\rm Pe}) = {\rm Pe}\,\exp({\rm Pe})\,{\rm E}_1({\rm Pe})$ with ${\rm E}_1({\rm Pe}) = \int_{\rm Pe}^\infty\exp(-t)/t \,dt$. Here, for consistency with PF simulations performed in two dimensions (2D), unless specified otherwise, we use a 2D version of the KGT model using $\kappa=1/R$ and ${\rm Iv}_{\rm 2D}({\rm Pe}) = \sqrt{\pi {\rm Pe}}\,\exp({\rm Pe})\,{\rm erfc}(\sqrt{\rm Pe})$ with $\text{erfc}(z) = 1 - (2/\sqrt{\pi}) \int_{0}^{z} e^{-t^{2}}\,dt$.
As expected and illustrated in Section~\ref{sec:results} (Fig.~\ref{fig:spacings}), the 2D KGT model results in higher undercooling at low velocity, but differences between 2D and 3D versions of the model vanish at high $V$.
This system of equations can be solved numerically, e.g. by fixing $V$ and solving iteratively for $(T,R)$.
The resulting $R(V) \to \infty$ when $V$ goes to either $V_c$ or $V_a$ \cite{kurz_theory_1986, rappaz_analysis_1990}, as the corresponding $T(V)$ tends toward $T_S(V)$.

\subsection{Phase-field modeling}
\label{method:phasefield}

\subsubsection{Model for rapid solidification}

We use a quantitative phase-field (PF) model for rapid solidification recently proposed by Ji and collaborators in \cite{ji_microstructural_2023,ji_quantitative_2024}. 
In this model, the spurious solute trapping effect associated with the use of a diffuse interface width $W>W_0$ (for computational convenience) is counterbalanced by an enhanced solute diffusivity through the interface. For a given scaling factor $S=W/W_0$, the diffusivity interpolation function across the interface can be selected to ensure that results for $S>1$ remain quantitatively close to those at $S=1$, particularly in terms of $k_V$ and $m_V$ across a wide range of $V$. 
By enabling quantitatively accurate simulations with $S > 1$, this approach allows the study of a solid-liquid interface morphology evolution at experimentally relevant length and time scales. 
For example, the model was used to successfully reproduce banding growth instability observed in {\it in situ} transmission electron microscopy (TEM) experiments of Al-Cu rapid solidification \cite{ji_microstructural_2023} and in additively manufactured (laser powder bed fusion) of a Mg alloy \cite{tourret_emergence_2024}. 

The evolution of the phase field $\phi$ and solute concentration field $c$ is described by the following equations \cite{ji_quantitative_2024}:
\begin{align} \label{phi_pfm}
\tau(\mathbf{n})\frac{\partial \phi}{\partial t} = ~&
\vec\nabla \cdot\left[ W(\mathbf{n})^2\vec\nabla \phi\right] +\phi-\phi^3 \nonumber\\
& +\sum_{\eta=x,y}\left[ \partial_\eta \left( |\vec\nabla\phi|^2 W(\mathbf{n}) \frac{\partial W(\mathbf{n})}{\partial (\partial_\eta\phi)}\right) \right] \nonumber\\
& -\lambda g'(\phi)\left[ c+\frac{T-T_{\rm M}}{m_{\rm e}} \exp\left\{b(1+g(\phi))\right\} \right] 
\end{align}
\begin{equation} \label{c_pfm}
\frac{\partial c}{\partial t} = \vec\nabla\cdot \left\{ D_{\rm L} q(\phi)c\vec\nabla[\ln c - bg(\phi)] \right\} ~,
\end{equation}
with $b=ln(k_e)/2$, $g(\phi) = 15/8(\phi - 2/3\phi^3 + 1/5\phi^5)$, and $\lambda = a^0_1 SW_0 \frac{b}{(k_e -1)} \frac{m_e}{\Gamma_{\rm sl}}$, where $a^0_1 = 2 \sqrt{2} /3$. 
Assuming a solute diffusivity $D_S = 0$ in the solid and $D=D_L$ in the liquid phase, an enhanced diffusivity is used across the interface, expressed as $D(\phi) = D_L q(\phi)$, with $q(\phi)$ defined as \cite{ji_quantitative_2024}:
\begin{equation} \label{q_phi_A}
    q(\phi) = A \frac{(1-\phi)}{2} - (A-1) \frac{(1-\phi)^2}{4}
\end{equation}
For a given $S=W/W_0 > 1$, the coefficient $A$ is determined following the approach proposed in \cite{ji_quantitative_2024} and summarized in \ref{appdx:convergence}, so as to minimize deviations in $k_V$ and $m_V$ from their $S=1$ values across the full velocity range. 

Anisotropic interface properties, namely excess free energy $\gamma({\bf n})=\gamma_0a_{\rm s}({\bf n})$ and kinetic coefficient $\mu_k({\bf n})=\mu_0a_{\rm k}({\bf n})$, are considered via anisotropy functions $a_{\rm s}({\bf n})$ and $a_{\rm k}({\bf n})$, where $\gamma_0$ and $\mu_0$ are the average values of $\gamma$ and $\mu_k$ and ${\bf n}=-\nabla\phi/|\nabla\phi|$ is the normal to solid-liquid interface.
The resulting anisotropic interface width and relaxation time are thus $W(\mathbf{n}) = SW_0a_{\rm s}({\bf n})$ and $\tau(\mathbf{n}) = \tau_0 a_{\rm s}({\bf n})^2/a_{\rm k}({\bf n})$, where $\tau_0 = (SW_0)^2/(\Gamma_{\rm sl} \mu_0)$. 
Here, we use the classical fourfold anisotropy functions $a_s(\theta)=1+\epsilon_s\cos(4\theta)$ and $a_k(\theta)=1+\epsilon_k\cos(4\theta)$, with $\theta=n_y/n_x=\tan^{-1}(\partial_y \phi / \partial_x \phi )$.

The classical frozen temperature approximation is considered, where a one-dimensional temperature field is imposed as 
\begin{equation} \label{FTA}
    T = T_0 + G(x-x_0-Vt)
\end{equation}
with $G$ the temperature gradient and $V$ is the steady-state growth velocity (i.e. the isotherms velocity, or pulling speed) with respect to a fixed reference temperature $T_0$ at a point $x_0$ at a time $t=0$. 

While the PF model is quantitative and its sensitivity to $(S,A)$ is subject to a careful convergence analysis (see \ref{appdx:convergence}), the exact $k_V$ and $m_V$ curves emerging from the PF model may deviate slightly from their classical CGM form in Eqs~\eqref{eq_kV}-\eqref{eq_mVme} \cite{ji_microstructural_2023,ji_quantitative_2024}.
The PF-consistent $k_V$ and $m_V$ can be calculated from the steady-state solution of the 1D coupled PF equations in a moving frame, as described in Appendix C of \cite{ji_quantitative_2024}.
Results of the KGT model, Eqs\,\eqref{KGT_main}-\eqref{KGTTipTemperature}, or $T_S(V)$ from Eq.~\eqref{tsolidus} can alternatively be calculated from the resulting $k_V$ and $m_V$ curves from these 1D PF calculations instead of Eqs~\eqref{eq_kV}-\eqref{eq_mVme}.

\subsubsection{Application to a Fe-Cr alloy}

Here, following the work of Pinomaa et al. \cite{pinomaa_significance_2020}, we focus on a binary Fe-Cr alloy as a surrogate for a 316L stainless steel in conditions relevant to AM via laser powder bed fusion (LPBF).

The rationale for selecting the Fe-Cr system is manifold. 
Rapidly solidified (e.g. welded, atomized, or additively manufactured) Fe alloys have – to the best of our knowledge – never been reported to exhibit banded microstructures.
This makes them a good candidate to address the open questions raised in Section~\ref{intro}, while leaving aside the complex dynamics of banding instability.
Stainless steel 316L is among the most technologically important steels with good printability via fusion-based AM processes.
This alloy also has a relatively low threshold for absolute stability ($V_a$) \cite{chadwick2021development}, which makes it susceptible to enter this rapid solidification regime for a relatively broad range of processes (e.g. even in direct energy deposition, typically operating at lower solidification rates than powder-bed fusion).
Moreover, previous results by Pinomaa and collaborators in \cite{pinomaa_significance_2020}, obtained with a different PF model, also provide a good baseline for comparison and benchmark of PF models for rapid solidification – a very active and relevant research field from both technological and scientific viewpoints.

The approximation of 316L by a pseudo-binary Fe-Cr is justified in \cite{pinomaa_significance_2020} (and references therein).
Essentially, in rapid solidification of 316L steel, the high-temperature BCC ferrite phase is suppressed, and the only significant phase transformation is from liquid to FCC. 
Due to the dominant role of Cr and the minimal impact of Ni and Mo on undercooling and diffusion, one may focus on Cr diffusion alone, treating nickel and molybdenum as passive elements that influence material parameters but do not actively affect solidification. 
Other elements (Mn, Si, N, P, S, C) are only present in very small proportions. 
Importantly, the resulting Fe-Cr pseudo-binary alloy, using material parameters summarized in Table~\ref{PFM_params} yields a freezing range  $\Delta T = |m_e | (1/k_e - 1)c_\infty = 15.67$\,K for a nominal concentration $c_\infty=17$\,wt\%Cr, which closely matches the CalPhaD calculated value of $15.1$\,K for the full 316L alloy (ThermoCalc, TCFE9 database) \cite{pinomaa_significance_2020}.

\begin{table*}[t!]
\centering
\begin{tabular}{| l | c | r | l |} 
\hline
  Parameter & Symbol & Value & Unit \\ \hline
  Solute (Cr) Nominal Concentration & $c_\infty$ & 15 -- 19 & wt\% \\
  Solvent (Fe) Melting Temperature & $T_M$ & 1811 & K \\
  Equilibrium Liquidus Slope & $m_e$ & $3.49$ & K/wt\% \\ 
  Equilibrium Partition Coefficient & $k_e$ & 0.791 & -- \\
  Gibbs Thomson Coefficient & $\Gamma_{\rm sl}$ & $3.47 \times 10^{-7}$ & K\:m \\
  Solute Diffusivity in Liquid & $D_L$ & $3.0 \times 10^{-9}$ & m$^2$/s \\
  Interface Kinetic Coefficient & $\mu_0$ & $0.3$ & m/s/K \\
  Kinetic Anisotropy Strength & $\epsilon_k$ & $0.13$ & -- \\
  Capillary Anisotropy Strength & $\epsilon_s$ & $0.018$ & -- \\
  Solid-Liquid Interface Thickness & $W_0$ & $1$ & nm \\
  Diffuse Interface Thickness & $W$ & 5 & $W_0$ \\
  Interface Diffusivity Coefficient & $A$ & 11 & -- \\
  Grid Spacing & $\Delta x$ & 0.8 & $W$ \\ \hline
\end{tabular}
\caption{Material and computational parameters used in phase-field simulation of pseudo-binary Fe-Cr approximation of 316L steel \cite{pinomaa_significance_2020}}
\label{PFM_params}
\end{table*}

\subsubsection{Numerical implementation}

Equations \eqref{phi_pfm} and \eqref{c_pfm} are solved using an explicit Euler time-stepping scheme with a finite difference spatial discretization. The implementation is written in C-based Compute Unified Device Architecture (CUDA) to enable massive parallelization across multiple Graphics Processing Units (GPUs). To reduce artificial grid anisotropy effects, the Laplacian and divergence terms in equations \eqref{phi_pfm} and \eqref{c_pfm} are discretized using operators that maintain isotropy up to order $\Delta x^2$ where $\Delta x$ is the grid spacing \cite{ji2022isotropic}. 
The most advanced, i.e. highest $T$, point along the solid-liquid interface is kept at a fixed location $x^*$ within the domain during the entire simulation by utilizing a moving frame in the $x$ direction, where $x^*$ is chosen far enough from the liquid boundary to allow a fully decaying concentration profile toward $c_\infty$ in cases with the highest solute partitioning (i.e. at lowest $V$). 
Further details about computational implementation are given in \cite{tourret_emergence_2024}.

Numerical parameters are listed at the bottom of Table \ref{PFM_params}. 
The upscaled diffuse interface thickness has $S=5$, and the corresponding coefficient $A=11$ was calculated by a 1D PF model analysis of $k_V$ and $m_V$ across the entire velocity range as described in \cite{ji_quantitative_2024} -- and shown in \ref{appdx:convergence} for the Fe-Cr system. 
The grid spacing $\Delta x=0.8W$ allows for an adequate spatial resolution of the $\phi$ profile across the interface. The explicit time step is selected based on the theoretical stability limit of the explicit scheme, considering both solute diffusivity and interface mobility (whichever is most restrictive), with a safety factor $R_t = 0.6$. 
These numerical parameters differ from those used by Pinomaa et al. \cite{pinomaa_significance_2020} since, in order to remain quantitative across the entire velocity range, we have to accommodate a much smaller value of $W=5~$nm, meaning more computationally demanding simulations ($W=25$~nm in \cite{pinomaa_significance_2020}). 

\subsubsection{Simulations}

Initial conditions for all PF simulations consist of a nearly planar interface located at the solidus temperature, with a small sinusoidal perturbation of amplitude $3\Delta x$ with 5 periods along the lateral ($y$) direction. The $\phi$ profile is initialized with the 1D stationary solution $\phi_0(x) = -\tanh[(x-x^*)/(\sqrt{2}W)]$ and the solute concentration field with the corresponding equilibrium $c(x) = c_\infty \exp\{b[g(\phi(x))+1]\}$ \cite{ji_quantitative_2024}. 
Boundary conditions are periodic in the lateral ($y$) directions and no-flux in the growth ($x$) direction. 

We explore a range of nominal concentration $c_\infty$ from 15 to 19\,wt\%\,Cr, slightly wider than the classical specification range for a 316L steel (16 to 18\,wt\%) in order to account for possible local solute segregation, e.g. due to fluid flow during solidification processing.
We consider a range of $G \in [10^5, 10^7]$\,K/m and $V\in [0.012, 0.6 ]$\,m/s, relevant to solidification under LPBF conditions, and also aiming to cover the cellular/dendritic transition toward a fully planar growth regime \cite{pinomaa_significance_2020}. 
Each PF simulation was run long enough to reach a microstructural steady-state with a well established microstructure for each value of $G$, $V$, and $c_\infty$. 

First, we perform a full factorial sampling with $c_\infty \in \{15,17,19\}$\,wt\%, $G \in \{10^5, 10^6, 10^7\}$\:K/m, and $V \in \{ 0.012$, 0.03, 0.06, 0.12, 0.3, $0.6 \}$\:m/s, with $S=5$ and $A=11$. 
For $G=10^7$\:K/m, PF simulations were completed in a (7.7\,\textmu m)$^2$ domain. For $G = 10^5$ and $10^6$\:K/m, the domain is (15.4\,\textmu m)$^2$ to accommodate for higher primary cell/dendrite spacings. 
In order to assess the effect of a wider diffuse interface width $W$, we also perform the 18 $(G,V)$ simulations for $c_\infty=17$\,wt\%, with $S=8$ and $A=18$.
Then, using the 54 simulations with $S=5$ and $A=11$ as initial training set, we use a Bayesian exploration strategy of the $(c_\infty,G,V)$ space, described later in Section~\ref{method:bayesian}, in order to pinpoint the location of the transition from cellular/dendritic and planar growth regimes.
Moreover, in order to explore a higher $G$ regime, close to its maximum limit beyond which a planar interface is expected to be stable across all velocities, we also perform, for $c_\infty=17$\,wt\%\,Cr, additional simulations at $G=2\times10^7$ and $4\times10^7$\:K/m for all six velocities listed above.

Each PF simulation is performed on one high-performance computing node. Each node is equipped with at least 2 Nvidia A100 GPUs and an Intel Xeon 8352Y (Ice Lake) CPU. The longest simulations (for $c_\infty=19$\,wt\%\,Cr, $G=10^5$~K/m, and $V=0.012$~m/s) lasted for 10 days on 8 Nvidia A100 GPUs on Texas A\&M University's high performance computer FASTER.

\subsection{Bayesian active learning with Gaussian processes}
\label{method:bayesian}

To accelerate the exploration of microstructures in Fe–Cr alloys in the rapid solidification regime relevant to the transition from cells/dendrites and planar interface (absolute stability), we adopt a Bayesian active learning framework that guides our design of experiments. 
The overall strategy is to construct a Gaussian process emulator (GPE) for the computationally expensive PF model described in Section~\ref{method:phasefield} and to iteratively improve this emulator by selecting new training points that are most informative about the decision boundary separating distinct solidification regimes.

\subsubsection{Gaussian process classification}

We model the relationship between input parameters -- nominal alloy composition (\( c_\infty \)), solidification velocity (\( V \)), and thermal gradient (\( G \)) -- and the resulting solidified morphology (cellular/dendritic or planar) using a Gaussian Process Classifier (GPC). The input space, denoted as \( \mathcal{X} \), spans \( c_\infty \in [0.15, 0.19] \), \( V \in [0.012, 0.6] \)\,m/s, and \( G \in [10^5, 10^7] \)\,K/m. To ensure adequate coverage of this space, we sample 50 evenly spaced points for \( c_\infty \) and \( V \), while \( G \) is sampled logarithmically over the same number of points.

\begin{figure*}[htb!]
  \centering
  \includegraphics[width=\textwidth]{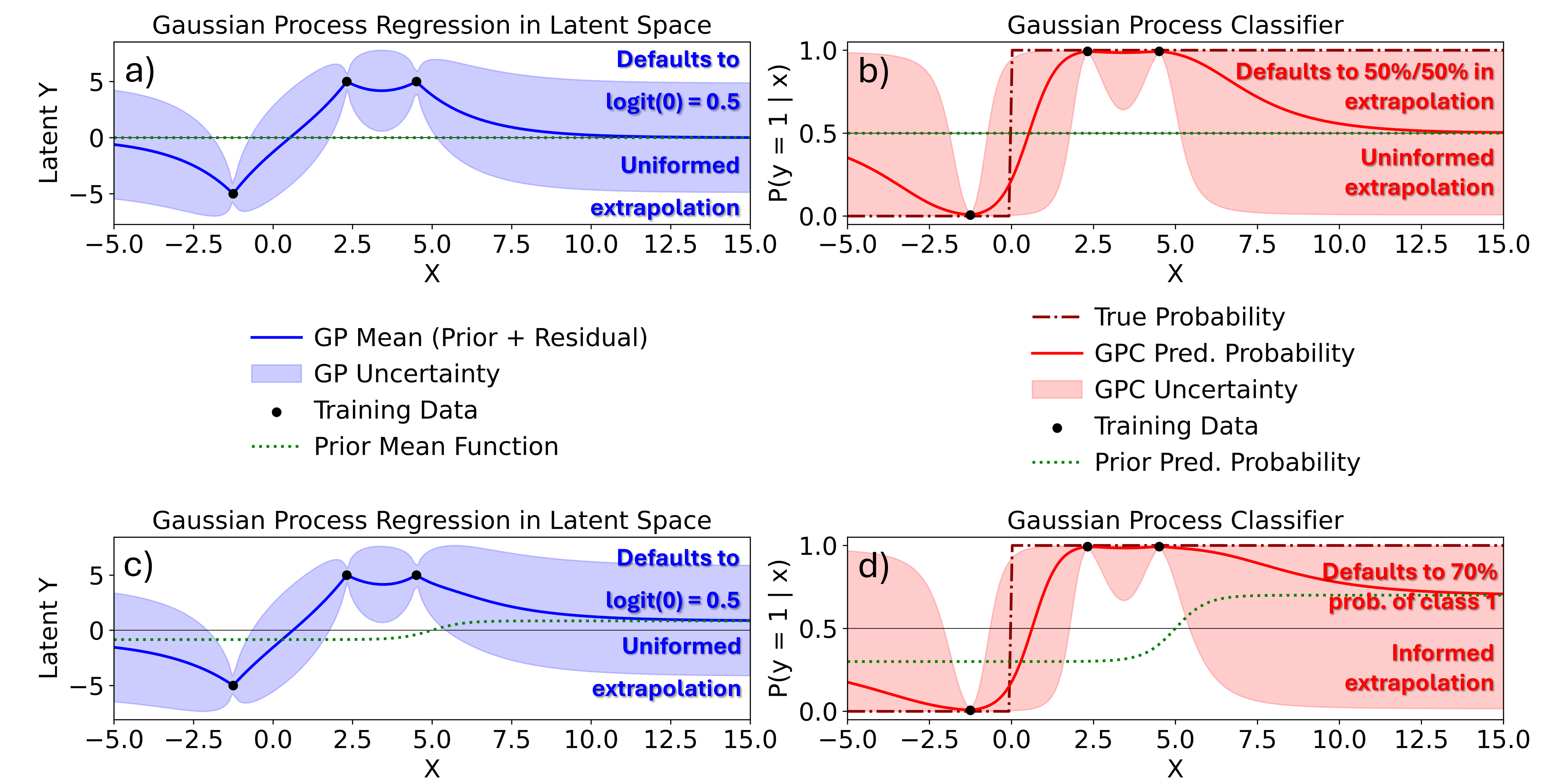}
  \caption{a) The latent GP trained on binary class observations. If a particular $x$ is observed to be class 1, the GP will pass through 5 at that $x$, and if observed to be class 0, the the GP will pass through -5. When extrapolating away from data, the latent GP defaults to an uninformed prior mean function of 0. b) The GP classifier providing a probabilistic classification prediction. The latent GP is passed through a sigmoid function in order to bound it between 0 and 1. When the GP classifier extrapolates away from data, the class predictions will tend toward a 50\% prediction for class 1 and a 50\% prediction for class 0. This prediction has the highest information entropy in a 2 class scenario. c) The latent GP with the informative prior mean function. When the latent GP begins predicting in regions without data, it will default to an informative prior mean function. In this illustrative case, the prior mean function corresponds to an intuition that at greater $x$ values, the true class values are more likely to be class 1. d) The GP classifier with an informative prior mean function. Once the GP classifier begins predicting away from training data, the class predictions at high values of $x$ tend toward 70\% probability of class 1.
  }\label{fig:GPCs}
\end{figure*}

Specifically, we assume that for any parameter set \( \mathbf{x} \in \mathcal{X} \), the probability of obtaining a planar morphology is given by:
\begin{equation}
    p(y = 1 \mid \mathbf{x}) = \sigma(f(\mathbf{x})),
\end{equation}
where \( y \in \{0, 1\} \) is a binary indicator of the solidified morphology, with \( y = 1 \) corresponding to planar morphology and \( y = 0 \) to cellular or dendritic morphology. The latent function \( f(\mathbf{x}) \) is modeled as a Gaussian Process:
\begin{equation}
    f(\mathbf{x}) \sim \mathcal{GP}(m(\mathbf{x}), k(\mathbf{x}, \mathbf{x}')).
\end{equation}
Here, $m(\mathbf{x})$ is the prior mean function and $k(\mathbf{x}, \mathbf{x}')$ is the kernel function, further discussed in Section \ref{PriorMeanFunc}, and \( \sigma(f) \) is a link function that maps the GP output to probabilities. 
In our case, we use the sigmoid function:

\begin{equation}
\sigma(f) = \frac{1}{1 + e^{-f}}
\end{equation}
This formulation allows us to probabilistically classify solidification outcomes while quantifying uncertainty, making it well-suited for active learning. Furthermore, GPCs can natively incorporate prior physics-based information, further accelerating the active learning process. A one-dimensional example of this is shown Figure \ref{fig:GPCs}. More details on Gaussian process classification with informative priors are provided in Ref \cite{hardcastle2025physics}.

\subsubsection{Prior mean function and kernel function of Gaussian process} \label{PriorMeanFunc}

A GP is defined by its prior mean function and its kernel function. Regarding the prior mean function, in this active learning scheme, the PF results are considered the ground truth for solidification morphology and the maximum value of the $T_S(V)$ curve from Eqs~\eqref{eq_kV}, \eqref{eq_mVme}, and \eqref{tsolidus} is used as the informative prior. Updating this prior with data via GP regression can be considered a Bayesian data-driven adjustment of the CGM using limited results from PF. This provides a refined estimate for dendritic/planar solidification boundary in \(\mathcal{X} = \big\{ (c_\infty, V, G) \mid c_\infty \in [0.15, 0.19], V \in [0.012, 0.6]\,{\rm m/s}, \, G \in [10^5, 10^7]\,{\rm K/m} \big\}\) based on both our prior knowledge and the observed data. In practice this means that in regions in $\mathcal{X}$ near where we have PF simulations, we will rely on that high-fidelity data. In regions in $\mathcal{X}$ where we do not have PF simulations, we will rely on the CGM. This corresponds to Figure \ref{fig:GPCs}c and d, where the latent GP and the GP classifier default to informative prior mean functions during extrapolation.

To achieve this, we modify the prior mean function of the latent GP before it is passed through the link function and becomes a GPC. An example of this is shown in Figure \ref{fig:GPCs}c. The posterior mean function (i.e. the trained GP) at any test input $\mu(\mathbf{x}_*)$ and the uncertainty in that prediction $\sigma^2(\mathbf{x}_*)$ are given by 
\begin{equation}
    \mu(\mathbf{x}_*) = m(\mathbf{x}_*) + k(\mathbf{x}_*, \mathbf{X}) \left[ K + \sigma_n^2 I \right]^{-1} \left( \mathbf{y} - m(\mathbf{X}) \right),
\end{equation}
\begin{equation}
\sigma^2(\mathbf{x}_*) = k(\mathbf{x}_*, \mathbf{x}_*) - k(\mathbf{x}_*, \mathbf{X}) \left[ K + \sigma_n^2 I \right]^{-1} k(\mathbf{X}, \mathbf{x}_*).
\end{equation}

In these expressions, $m(\mathbf{x}_*)$ is the prior mean function evaluated at the test input $\mathbf{x}_*$, representing our initial belief before observing any data. In the case of GPCs, we set $m(\mathbf{x}_*) = \text{logit}(p_p)$, where $p_p$ is the prior probability of a class label from the prior model. By setting $m(\mathbf{x}_*) = \text{logit}(p_p)$ this ensures that when the latent GP is passed through the sigmoid link function, the prior probability of the resultant GPC will be $p_p$ (recall that logit is the inverse of a sigmoid). In Figure \ref{fig:GPCs}d, this is $p_p = 0.7$. Finally, $m(\mathbf{X})$ is the vector of prior mean function values evaluated at all training inputs. 

Regarding terms related to the kernel function, $k(\mathbf{x}_*, \mathbf{X})$ is a vector of covariance (kernel) values between the test input $\mathbf{x}_*$ and each of the training inputs in $\mathbf{X}$, defining their correlation. $K=K(\mathbf{X}, \mathbf{X})$ is the covariance matrix of all training points, where each entry is given by $k(\mathbf{x}_i, \mathbf{x}_j)$. $\sigma_n^2 I$ accounts for the noise in the training observations, where $\sigma_n^2$ is the noise variance and $I$ is the identity matrix. $\mathbf{y}$ is the vector of observed outputs for the training points $\mathbf{X}$.

In practice, to define the prior class probabilities, we queried CGM predictions across all values that we sampled in $\mathcal{X}$. This resulted in 125000 combinations of $c_\infty$, $G$, and $V$. Because the CGM is a fast acting analytical model, this query was completed with negligible time in a computationally inexpensive fashion. This is an important aspect about the CGM-based criterion that makes it viable as an informative prior.

The kernel selected for this study is the Matérn kernel, widely favored for modeling functions that exhibit varying degrees of smoothness \cite{rasmussen2006gaussian}. The commonly utilized formulation of the Matérn kernel is given by:
\begin{equation}
    k(x, x') = \frac{2^{1 - \nu}}{\Gamma(\nu)} \left( \frac{\sqrt{2\nu}|x - x'|}{\ell} \right)^{\nu} K_{\nu} \left( \frac{\sqrt{2\nu}|x - x'|}{\ell} \right),
\end{equation}
where \( \nu > 0 \) is a parameter governing the smoothness of the function, \( \ell \) is the length scale parameter, \( \Gamma \) denotes the gamma function, and \( K_{\nu} \) represents the modified Bessel function of the second kind. In this work, the smoothness parameter was set to \( \nu = 0.5 \). Normally, the length scale (\( \ell \)) for each dimension is optimized within the range from 0.05 to 1 by maximizing the log marginal likelihood: 
\begin{align}
\log L(\mathbf{y}|\mathbf{X}, \ell) = & -\frac{\mathbf{y}^\top \left[ K + \sigma_n^2 I \right]^{-1}\mathbf{y}}{2} \nonumber \\
& - \frac{\log\left|\left[ K + \sigma_n^2 I \right]\right|}{2} - \frac{n\log(2\pi)}{2},
\label{eq:log_marginal_likelihood}
\end{align}
where $n$ is the number of observations.
For iterations 1 through 4, the log marginal likelihood was used to tune hyperparameters. In the final iteration, in order to prevent overfitting and to yield smoother boundaries in the final probability map, the length scales were fixed at \( \ell_V=0.18 \), \( \ell_G=0.4 \), and \( \ell_{c_\infty}=0.3 \). 

\subsubsection{Active learning initialization and classification} \label{ActivLearnInit}

In addition to the prior given by the maximum of the $T_S(V)$ curve from Eq.\,\eqref{tsolidus} considering Eqs\,\eqref{eq_kV}-\eqref{eq_mVme} (see Section~\ref{PriorMeanFunc}), Bayesian active learning requires a 0$^{\rm th}$ ``seed'' iteration to initialize the closed-loop search. 
For this, we use a grid of 54 points $\mathbf{x_0} = ({\text{c}_{\infty}}_0, \text{G}_0, \text{V}_0) \in \mathcal{X}$, where ${\text{c}_{\infty}}_0 \in \{15$, 17, 19\}\,wt.\%Cr, G$_0 \in \{10^5$, $10^6$, $10^7$\}\,K/m and V$_0 \in 
\{0.012$, 0.03, 0.06, 0.12, 0.3, 0.6\}\,m/s as the initial training set. 
The PF simulation results for these initial points are presented later in Figures~\ref{fig:C15grid}, \ref{fig:C17grid}, and \ref{fig:C19grid}. 

Each simulation was labeled as ``planar'', ``intermediate'', or ``cellular/dendritic'', which in terms of planar probability $p(y = 1 \mid \mathbf{x_0})$ are equal to 1.0, 0.5, and 0.0, respectively. 

The ``cellular/dendritic'' class was used for regions marked with a clear, measurable, and steady primary arm spacing, regardless of whether the pattern exhibits more cellular or dendritic features. 
Here, for a given simulation, we measured the primary dendrite arm spacing (PDAS) using the concentration profile $c(y)$ along a line at $(x=x_1)$, with $x_1$ located behind the solidification front (by $\approx10\%$ of the solid domain length in $x$) and counting the number of local peaks therein.
We found that this method was robust to identify interdendritic regions even for shallow cells, as they leave a clear solute segregation trace even far behind the solidification front (see concentration maps in later Figures~\ref{fig:C15grid}-\ref{fig:C19grid}).

Patterns classified as ``planar'' were straightforward to ascertain from simple visual observation, but they could be also automatically classified due to their flat $c(x_1,y)$ concentration profile, or by measuring the difference between the minimum and maximum $x$-position of the solid-liquid interface.

The ``intermediate'' classification was used when the resulting pattern was not planar, but also not clearly marked with an unambiguous, measurable, and steady primary spacing (see examples in Section~\ref{sec:results} and discussion in Section~\ref{sec:disc}).
Representing the intermediate class with a probability of $p = 0.5$ relies on the assumption that when the PF model predicts intermediate solidification behavior, it seems reasonable to assume an equal 50\% probability of planar or cellular/dendritic solidification. This assumption also simplifies the active learning scheme by reducing it to a binary classification problem.

\subsubsection{Active learning acquisition function} \label{aqf_section}

Once an initial Gaussian Process Emulator (GPE) is built from the set of training points described in the previous section, we iteratively select the next PF simulations to run by maximizing an acquisition function. Our active learning objective is to efficiently identify the decision boundary between different solidification regimes. To do so, we define an acquisition function \(a(\mathbf{x})\) that is a function of the GP’s uncertainty in its class label predictions.

Because of the computational cost of PF simulations, sequential active learning -- i.e., conducting one simulation at a time and waiting for the result before recommending the next -- is not viable. 
Instead, the active learning scheme must operate in batch: at each iteration, a batch of multiple PF simulations is selected and run in parallel on different points \(\mathbf{x} \in \mathcal{X}\). 

To achieve this, we use a heuristic but practical approach. Specifically, we begin by computing the Shannon entropy, $H(p(\mathbf{x}))$,  of the predicted class probabilities at all candidate points in \(\mathcal{X}\). For a binary classification problem, let \(p(\mathbf{x})\) denote the predicted probability for one class (and hence \(1 - p(\mathbf{x})\) for the other). The Shannon entropy is computed as:
\begin{equation}
H(p(\mathbf{x})) = -\left[ p(\mathbf{x}) \log p(\mathbf{x}) + \bigl(1 - p(\mathbf{x})\bigr) \log \bigl(1 - p(\mathbf{x})\bigr) \right].    
\end{equation}
We then select the top 2.5\% of points in \(\mathcal{X}\) with the highest entropy values, denoting this subset as \(\mathcal{X}_{\text{high}}\). These points correspond to regions of maximum uncertainty, typically near the decision boundary between solidification regimes. 

To reduce redundancy and promote spatial diversity, we cluster the points in \(\mathcal{X}_{\text{high}}\) into 10 groups using the KMedoids algorithm. 
KMedoids is a robust clustering method that selects actual data points as centers and its sensitivity to outliers is low \cite{park2009simple}. The medoid of each cluster -- i.e. the point minimizing the total distance to others in its cluster -- is selected as a candidate for the next simulation. Let \(\mathcal{X}_{\text{medoid}}\) denote this set of 10 selected medoids.
The acquisition function \( a(\mathbf{x}) \) is then defined as an indicator function:
\begin{equation}
a(\mathbf{x}) =
\begin{cases}
1 & \text{if } \mathbf{x} \in \mathcal{X}_{\text{medoid}}, \\
0 & \text{otherwise}.
\end{cases}
\end{equation}
This ensures that only the 10 spatially diverse, high-uncertainty medoids are chosen as candidates for the next batch of simulations. 

In summary, using an initial prior with a cellular-planar threshold given by the maximum of the $T_S(V)$ curve from Eq.\,\eqref{tsolidus} (Section~\ref{PriorMeanFunc}) and an initial mapping $\mathbf{x_0}$ from 54 PF simulations (Section~\ref{ActivLearnInit}), we perform 5 iterations of 10 PF simulations.
At each iteration, the next 10 $(c_\infty,G,V)$ configurations are selected via the active learning approach described above in order to maximize the acquisition function (i.e. focusing on regions with greater uncertainty in the cellular/planar classification), while ensuring spatial diversity of the next cluster of points.

\section{Results} \label{sec:results}

\begin{figure*}[t!]
  \centering
  \includegraphics[width=\textwidth]{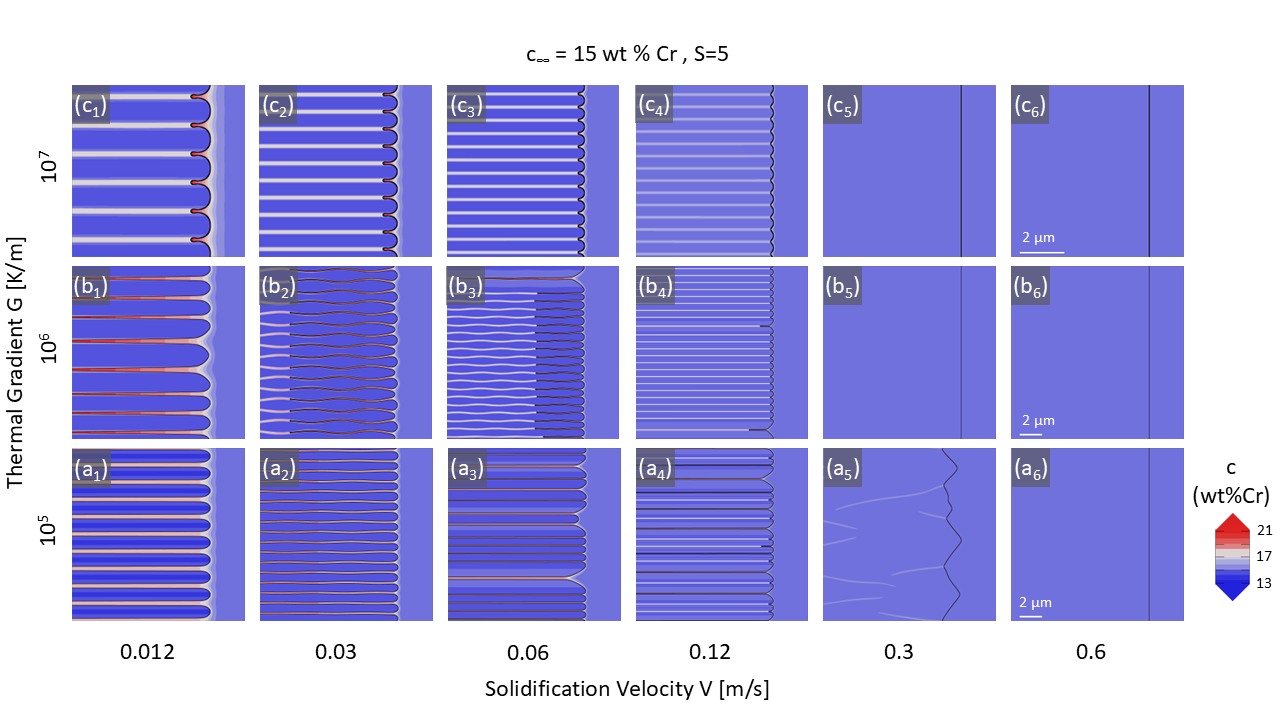}
  \caption{Final state of solute concentration field (color maps) and solid-liquid interface (black line) for various ($G, V$) conditions at a nominal concentration $c_\infty = 15$\,wt\%\,Cr from PF simulations with $S=5$, $A=11$.
}\label{fig:C15grid}
\end{figure*}

\begin{figure*}[t!]
\centering
\includegraphics[width=\textwidth]{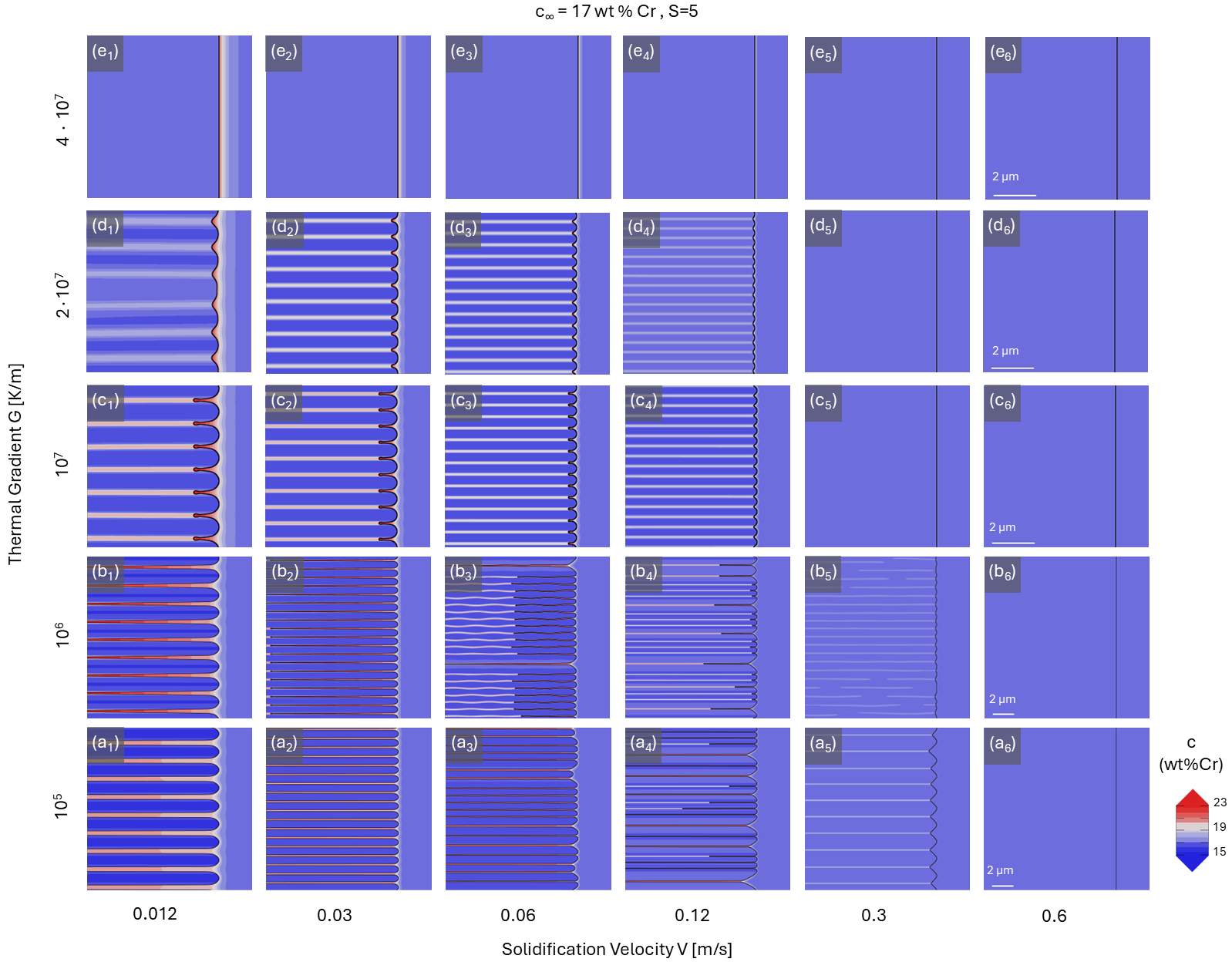}
  \caption{Final state of solute concentration field (color maps) and solid-liquid interface (black line) for various ($G, V$) conditions at a nominal concentration $c_\infty = 17$\,wt\%\,Cr from PF simulations with $S=5$, $A=11$.
}
\label{fig:C17grid}
\end{figure*}

\begin{figure*}[t!]
\centering
\includegraphics[width=\textwidth]{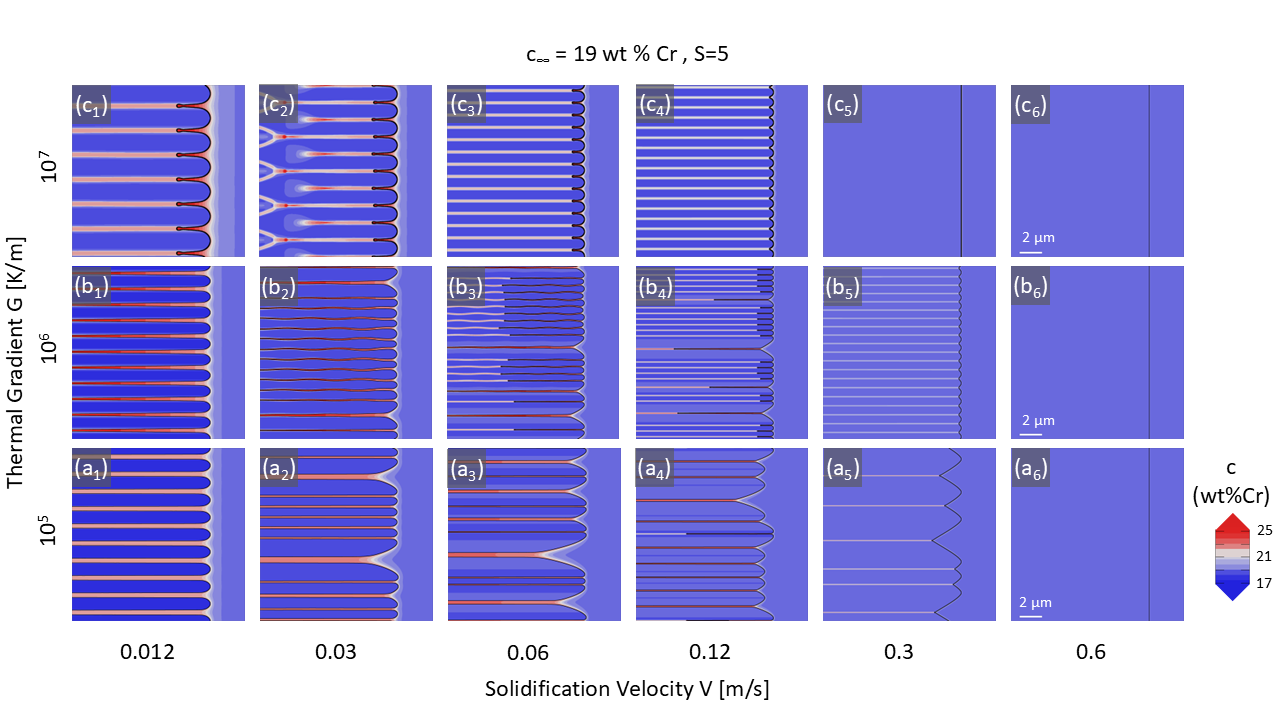}
  \caption{Final state of solute concentration field (color maps) and solid-liquid interface (black line) for various ($G, V$) conditions at a nominal concentration $c_\infty = 19$\,wt\%\,Cr from PF simulations with $S=5$, $A=11$.\\
}
\label{fig:C19grid}
\end{figure*}

\begin{figure*}[t!]
  \centering
  \includegraphics[width=\textwidth]{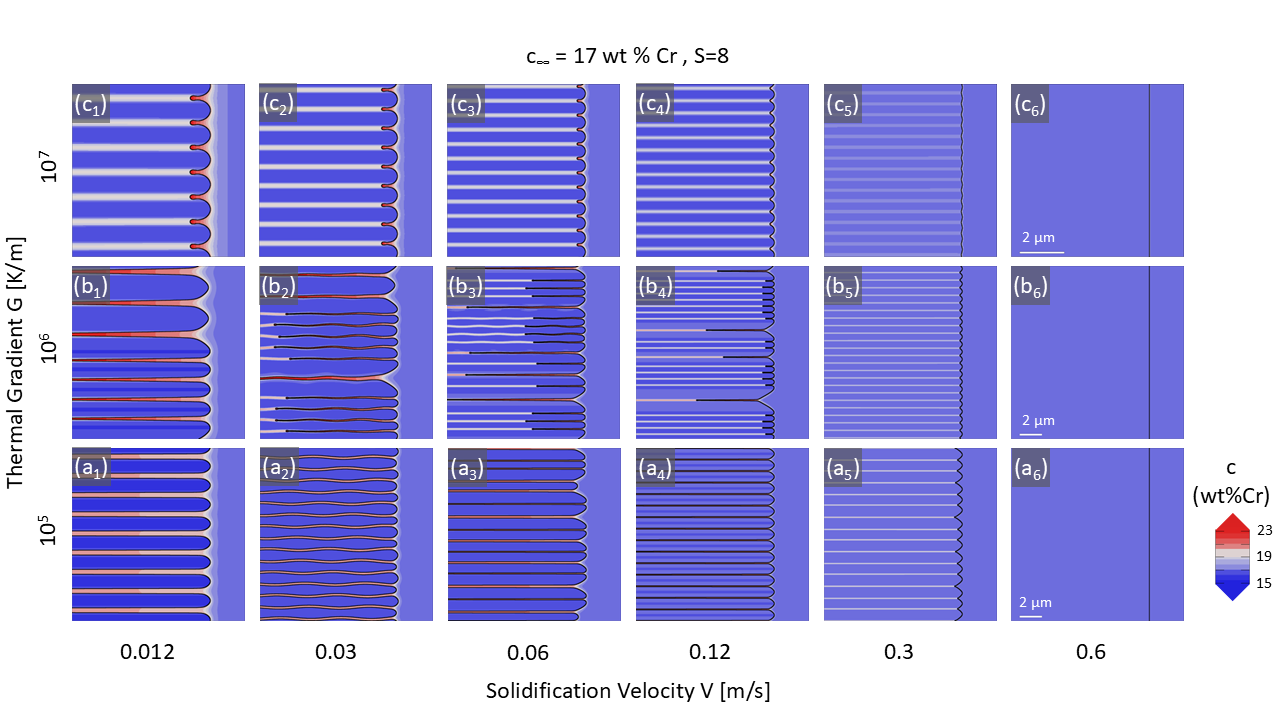}
  \caption{Final state of solute concentration field (color maps) and solid-liquid interface (black line) for various ($G, V$) conditions at a nominal concentration $c_\infty = 17$\,wt\%\,Cr from PF simulations with $S=8$, $A=18$.
}\label{fig:S8grid}
\end{figure*}

Figure~\ref{fig:C15grid}, \ref{fig:C17grid}, and \ref{fig:C19grid} show the final microstructures for the full $(G,V)$ map of completed PF simulations with $c_\infty = 15$, 17, and 19\,wt\% Cr, respectively, using $S=5$ and $A=11$. 
The color map shows the Cr concentration and black lines representing the solid-liquid interface ($\phi = 0$). 
The growth direction ($x+$) is left to right. 
The combination of these three figures includes the initial parameter set used in the Bayesian exploration of the cellular/planar transition in the $(c_\infty,G,V)$ space, as well as, for $c_\infty = 17$\,wt\% Cr (Fig.~\ref{fig:C17grid}), the additional high $G$ values explored until full planar stability. 
Figure \ref{fig:S8grid} shows a similar map for $c_\infty = 17$\,wt\%\,Cr, but using $S=8$ and $A=18$, which allows us to compare how larger $S$ values can affect the model predictions.


\begin{figure}[htbp!]
\centering
\includegraphics[width=.95\columnwidth]{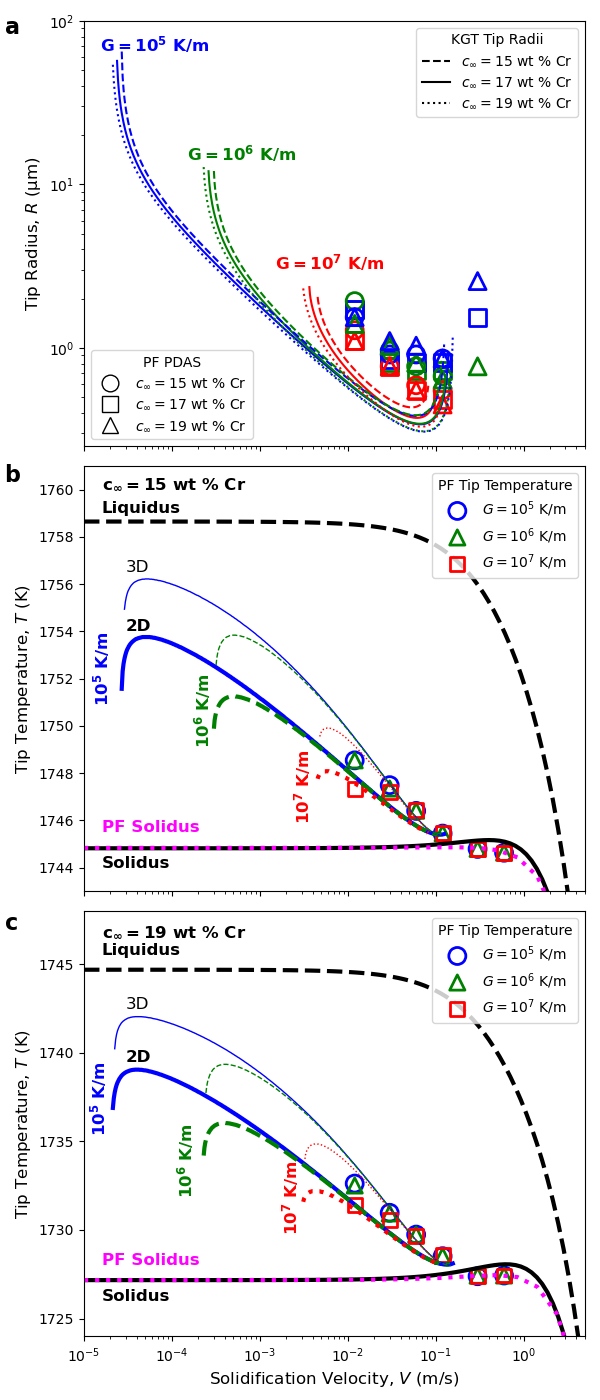}
\caption{
Comparison of PF predictions (symbols) with classical theories (lines) for microstructural length scales (a) and growth temperature (b,c) as a function of growth velocity $V$. 
(a) Dendrite tip radius calculated from KGT model, Eqs.~\eqref{KGT_main}-\eqref{KGTTipTemperature}, compared to primary dendritic arm spacing (PDAS) from PF simulations. 
(b,c) KGT model prediction of dendrite tip temperature for different $G$ (blue, green, and red lines), solidus and liquidus temperatures from Eqs~\eqref{tsolidus}-\eqref{tliquidus} (black lines), and solidus temperature from Eq.~\eqref{tsolidus} using $k_V$ and $m_V$ from 1D PF calculations (pink dotted line), compared to PF tip temperature (symbols) for $c_\infty = $ (b) 15 and (c) 19\,wt.\%\,Cr.
Both 2D (thick lines) and 3D (thin lines) versions of the KGT model predictions are shown in (b) and (c).
Only 2D predictions appear in (a) as they are barely distinguishable from their 3D equivalent.
}
\label{fig:spacings}
\end{figure}

Figure~\ref{fig:spacings} presents a comparison between classical theories (lines) -- namely the KGT model predictions from Eqs\,\eqref{KGT_main}-\eqref{KGTTipTemperature} as well as $T_L(V)$ and $T_S(V)$ from Eqs\,\eqref{tliquidus}-\eqref{tsolidus} -- and PF results for $S=5$, $A=11$ (symbols).
Fig.~\ref{fig:spacings}(a) actually compares dendrite tip radius from the KGT model with primary spacings from PF simulations for all cases, while Fig.~\ref{fig:spacings}(b)-(c) shows the PF-predicted highest solid-liquid interface temperature (i.e. either cell/dendrite tip or planar interface temperature) for $c_\infty=$ (b) 15 or (c) 19\,wt\%\,Cr.
The KGT-predicted tip temperatures are shown for both 2D (thick lines) and 3D (thin lines) versions of the model (see Section~\ref{sec:rapid}).
Fig.~\ref{fig:spacings}(b)-(c) also include the $T_S(V)$ line calculated from Eq.\,\eqref{tsolidus} but using $k_V$ and $m_V$ calculated from PF calculations instead of Eqs~\eqref{eq_kV}-\eqref{eq_mVme}, as presented in \cite{ji_quantitative_2024} (Appendix C therein).

\begin{figure}[t!]
\centering
\includegraphics[width=\columnwidth]{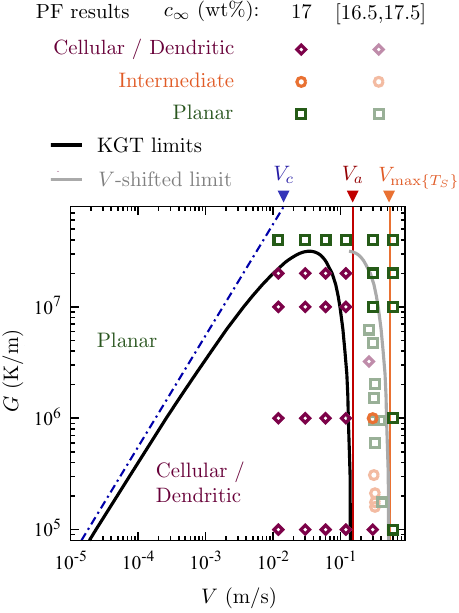}
\caption{
Map comparing classical theories (lines) and PF results (symbols) in the $(G,V)$ space for  $c_\infty\approx17$\,wt\%\,Cr.
PF results use darker symbols for calculations at exactly $c_\infty=17$\,wt\%\,Cr (Fig.~\ref{fig:C17grid}) and lighter colors for simulations at $c_\infty\in[16.5,17.5]$\,wt\%\,Cr performed during the Bayesian-guided exploration.
Analytical theories include planar stability limits from the KGT model (black line), constitutional undercooling velocity $V_c$ from Eq.~\eqref{eq:const_underc} (blue dash-dotted line), absolute stability $V_a$ from Eq.~\eqref{eq:absolutestability} (red vertical line) or using the maximum of $T_S(V)$ from Eq.~\eqref{tsolidus}, denoted $V_{\max\{T_S\}}$ (orange vertical line).
The plot also reproduces the high-$V$ KGT-predicted limit (see Section~\ref{disc:pf_vs_kgt}) shifted by a factor $V_{\max\{T_S\}}/V_a$ (gray line).
}
\label{fig:gv}
\end{figure}

PF results and classical theories are further compared, for $c_\infty\approx17$\,wt\%\,Cr, in Figure~\ref{fig:gv}.
Therein, in a $(G,V)$ map similar to Fig.~\ref{fig:gv_schem}, we plot the KGT predictions of the planar stability limits (black line), compared to results of PF simulations (symbols), and analytically calculated values of $V_c$ and $V_a$ (see Section~\ref{sec:rapid}).
The initial mapping for $c_\infty=17$\,wt\%\,Cr (Fig.~\ref{fig:C17grid}) appears as darker symbols, while lighter color shades are used for simulations performed during the Bayesian exploration within a range $c_\infty\in[16.5,17.5]$\,wt\%\,Cr.

\begin{figure*}[htbp!]
\centering
\includegraphics[width=\textwidth]{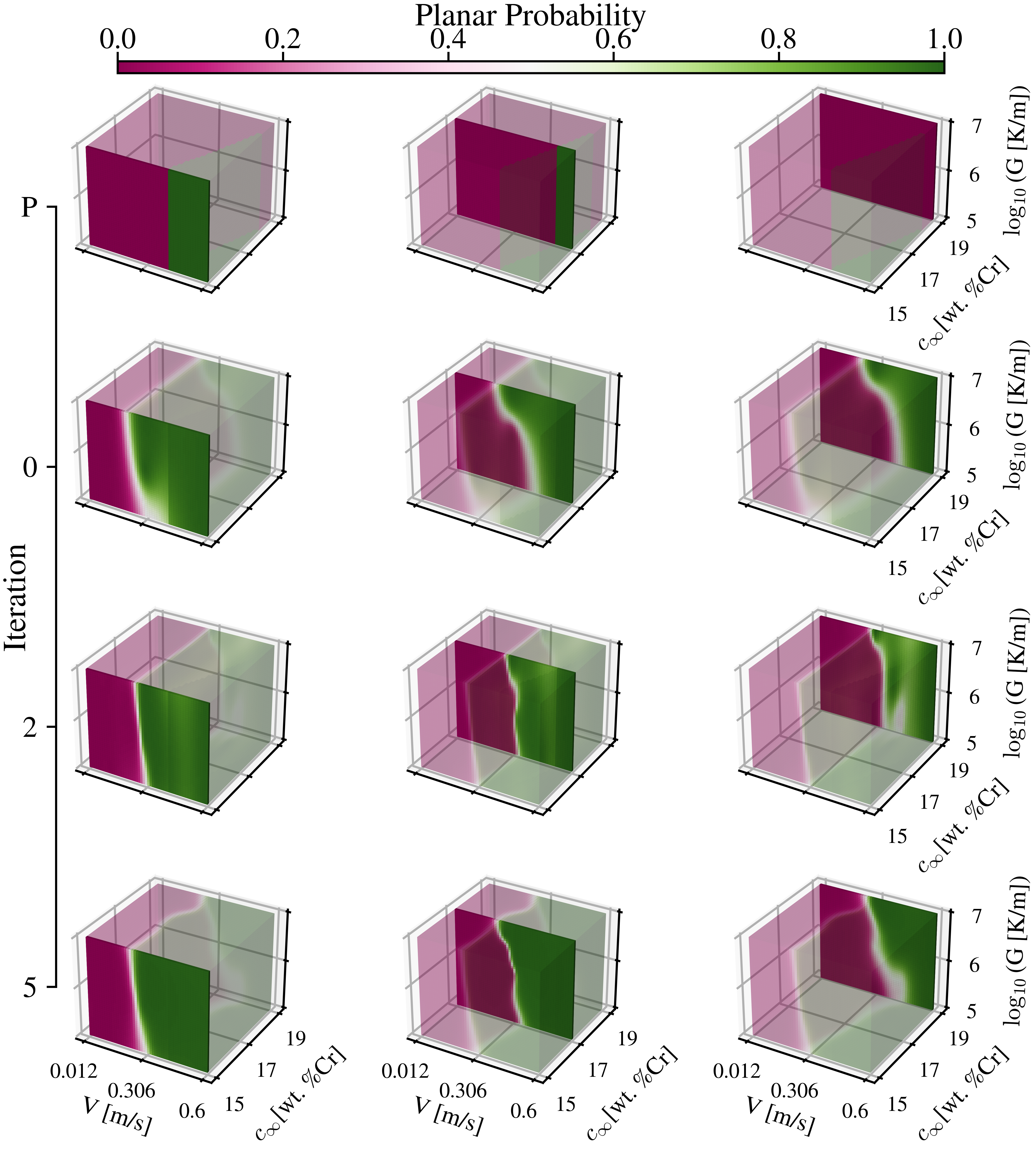}
\caption{
Evolution of the microstructure selection map in the $(G,V,c_\infty)$ space, exhibiting planar ($p=1$, green), cellular ($p=0$, purple), and intermediate ($p=0.5$, white) regions, illustrating the CGM-based prior (P), the full-factorial PF-based iteration~0 (Figures \ref{fig:C15grid}-\ref{fig:C19grid}), and two iterations of the Bayesian exploration process (rows), highlighting $c_\infty =$ 15, 17, and 19\,wt\%\,Cr planes (columns).
}
\label{fig:iterations_grid}
\end{figure*}

Figures~\ref{fig:cellplanarspace} and \ref{fig:iterations_grid} show the results of the GP classification procedure across active learning iterations.
Fig.~\ref{fig:cellplanarspace} shows the final selection map after 5 active learning iterations in the $(c_\infty,G,V)$ space, highlighting the $c_\infty=$ (a) 15, (b) 17, and (c) 19\,wt\%\,Cr planes, as well as characteristic microstructures corresponding to configurations summarized in Table~\ref{InsetsFigure}.
Figure \ref{fig:iterations_grid} shows the evolution of the predicted microstructure selection map made by the GP model at different iterations of the exploration process, including the CGM-based prior (P), the PF-based full-factorial ``seed'' iteration (0) and two of the five successive updates.

The joint Supplementary Material contains detailed $(G,V,c_\infty)$ parameters, as well as primary spacings, and classification ({\it cellular}/{\it dendritic}, {\it intermediate}, or {\it planar}) for every simulations presented in this article, including Bayesian exploration iterations and corresponding final microstructures. Videos corresponding to panels (c$_1$)-(c$_4$) of Figure~\ref{fig:cellplanarspace} are also attached.

\begin{figure*}[htbp!]
\centering
\includegraphics[width=\textwidth]{Fig9_FinalProbabilityMap.png} 
\caption{
Microstructure selection map in the $(G,V,c_\infty)$ space from the Gaussian process map, exhibiting planar ($p=1$, green), cellular ($p=0$, purple), and intermediate ($p=0.5$, white) regions, highlighting (a) 15, (b) 17, and (c) 19\,wt\%\,Cr planes. 
Microstructures are illustrated for data points as listed in Table~\ref{InsetsFigure}. 
All scale bars are of size 2\,\textmu m. Videos for panels (c$_1$) to (c$_4$) are included as supplementary material. 
}
\label{fig:cellplanarspace}
\end{figure*}

\begin{table}[t!]
\centering
\begin{tabular}{| c | c | c | c |} 
\hline
  Label & $G$  & $V$  & $c_\infty$  \\ 
  \hline
  a$_1$ & $3.728\times10^5$ & 0.276 & 15.408 \\ 
  a$_2$ & $10^6$ & 0.600 & 15.000 \\ 
  a$_3$ & $10^5$ & 0.030 & 15.000 \\ 
  a$_4$ & $2.812\times10^5$ & 0.300 & 15.490 \\ 
  \hline
  b$_1$ & $3.089\times10^5$ & 0.312 & 17.449 \\ 
  b$_2$ & $10^7$ & 0.600 & 17.000 \\ 
  b$_3$ & $10^5$ & 0.060 & 17.000 \\ 
  b$_4$ & $2.121\times10^5$ & 0.324 & 17.040 \\ 
  \hline
  c$_1$ & $7.197\times10^5$ & 0.348 & 18.510 \\ 
  c$_2$ & $10^6$ & 0.600 & 19.000 \\ 
  c$_3$ & $10^5$ & 0.120 & 19.000 \\ 
  c$_4$ & $3.089\times10^5$ & 0.444 & 18.755 \\ 
  \hline
   Units & K/m & m/s & wt\%\,Cr \\ 
\hline
\end{tabular}
\caption{
Conditions for PF simulated microstructures illustrated in Fig.~\ref{fig:cellplanarspace}.
}
\label{InsetsFigure}
\end{table}

\section{Discussion} \label{sec:disc}

\subsection{Microstructure selection}

Figures~\ref{fig:C15grid}, \ref{fig:C17grid}, and \ref{fig:C19grid}, combined with Figs~\ref{fig:spacings} and \ref{fig:gv}, give first idea of the effect of the three main control parameters $(G,V,c_\infty)$ on the selected rapid solidification microstructure.

\subsubsection{Comparison with classical theories} \label{disc:pf_vs_kgt}

The limit of planar stability at high $V$ and high $G$ predicted by the PF model shows interesting similarities with classical theories (Fig.~\ref{fig:gv}).
The maximum $G$ for non-planar growth from KGT calculations, $G_{\rm max}\approx3.16\times10^7$~K/m (or $G_{\rm max}\approx3.00\times10^7$~K/m using the 3D version of the model), is in excellent agreement with PF results, which show cells/dendrites at $G=2\times10^7$~K/m, but only planar interfaces at $G=4\times10^7$~K/m. 
The decrease of the absolute stability threshold velocity when $G\to G_{\rm max}$ from the KGT model matches PF results (as discussed later in Section~\ref{disc:G}), in contrast to $V_a=0.154$~m/s from Eq.~\eqref{eq:absolutestability} or $V_{\max\{T_S\}}=0.534$~m/s from Eq.~\eqref{tsolidus} that do not have any dependence upon $G$.
At low $G$, the restabilization of a planar interface occurs closer to $V_{\max\{T_S\}}$ than $V_a$, which is consistent with recent experimental observations of dendrites growing at $V_a<V<V_{\max\{T_S\}}$ in a laser-melted ternary Ni alloy \cite{tourret2023morphological}.
Hence, shifting the KGT-predicted absolute stability limit by a factor $V_{\max\{T_S\}}/V_a$, as pictured in Fig.~\ref{fig:gv} (gray solid line), results in a reasonable -- though slightly overestimated -- approximation of the PF absolute stability limit including its dependence on $G$.
While approximate, this shifted limit is advantageous as it can be calculated easily from a simple KGT calculation (no costly PF calculation required).

In terms of operating tip temperature (Fig.~\ref{fig:spacings}b-c), PF results appear in good agreement with classical theories, with the front temperatures smoothly transitioning from the KGT-predicted $T(V)$ in the cellular/dendritic regime to the $T_S(V)$ curve in the planar growth regime.
The only notable deviations appear for the highest $G$, which, as discussed below (Section~\ref{disc:G}), leads to a cellular interface pattern that is not appropriately represented by the KGT model assumptions.
Interestingly, the 2D PF results appear closer to the 3D KGT prediction but, as expected, the difference between 2D and 3D versions of the KGT model essentially vanishes at high $V$. 

\clearpage
\subsubsection{Effect of the alloy concentration $c_\infty$} \label{disc:conc}

Due to the relatively narrow range of $c_\infty$ explored here, its effect appears minor compared to that of $G$ and $V$.
Still, a lower concentration appears to lower the absolute stability velocity threshold, as seen for $V=0.3$~m/s and $G=10^6$~K/m, i.e. panel (b$_5$) in Figs~\ref{fig:C15grid}-\ref{fig:C19grid}.

The dependence of $V_a$ upon $c_\infty$ alone appears weaker than predicted using Eq.~\eqref{eq:absolutestability} or $V_{\max\{T_S\}}$ from Eq.~\eqref{tsolidus} for the morphological transition threshold.
Indeed, Eq.~\eqref{eq:absolutestability} results in a variation of $V_a$ from 0.131\,m/s for $c_\infty=15$\,wt\% up to 0.162\,m/s for $c_\infty=19$\,wt\%.
The maximum of $T_S(V)$ from Eq.~\eqref{tsolidus} changes from 0.405\,m/s for $c_\infty=15$\,wt\% to 0.626\,m/s for $c_\infty=19$\,wt\%.
Meanwhile, focusing on $G=10^7$~K/m as a reference, PF results (Fig.~\ref{fig:cellplanarspace}) show that this transition shifts from 0.222\,m/s ($c_\infty=15$\,wt\%) to 0.246\,m/s (19\,wt\%).
Lower $G$ regions are more complex as they exhibit a broader ``intermediate'' region (see Section~\ref{disc:G}).

\subsubsection{Effect of the growth velocity $V$}

As expected, the main effect of increasing $V$ is to decrease the primary dendrite (or cell) arm spacing (Fig.~\ref{fig:spacings}a) and ultimately to cause a transition to a fully planar microstructure with full solute trapping at high $V$ (Fig~\ref{fig:C15grid}-\ref{fig:C19grid}).
As mentioned in Section~\ref{disc:pf_vs_kgt}, the absolute stability threshold from PF simulations is close to the maximum of the $T_S(V)$ curve (Fig.~\ref{fig:spacings}b-c).
This $V_{\max\{T_S\}}$ is even higher when considering the PF-calculated $k_V$ and $m_V$ (pink dotted lines).
Interestingly, in most cases, the primary spacing exhibits a notable increase just before the transition to a planar front as $V \to V_a$ (Fig.~\ref{fig:spacings}a).
This is most visible, for instance, in panel (a$_5$) of Figs~\ref{fig:C17grid} and \ref{fig:C19grid}.
This could be a general behavior, at least for relatively low $G$.
Hence, in Fig.~\ref{fig:spacings}, this PDAS behavior emulates the KGT prediction with a sharp increase in $R$ when $V\to V_a$, however with a slightly higher $V_a$ closer to $V_{\max\{T_S\}}$ for the PF results.
Note that this increase of spacing as $V\to V_a$ is not observed for the highest $G$, which results in cellular patterns rather than dendrites, as discussed in Section~\ref{disc:G}.

\subsubsection{Effect of the temperature gradient $G$} \label{disc:G}

\begin{figure}[b!]
\centering
\includegraphics[width=\columnwidth]{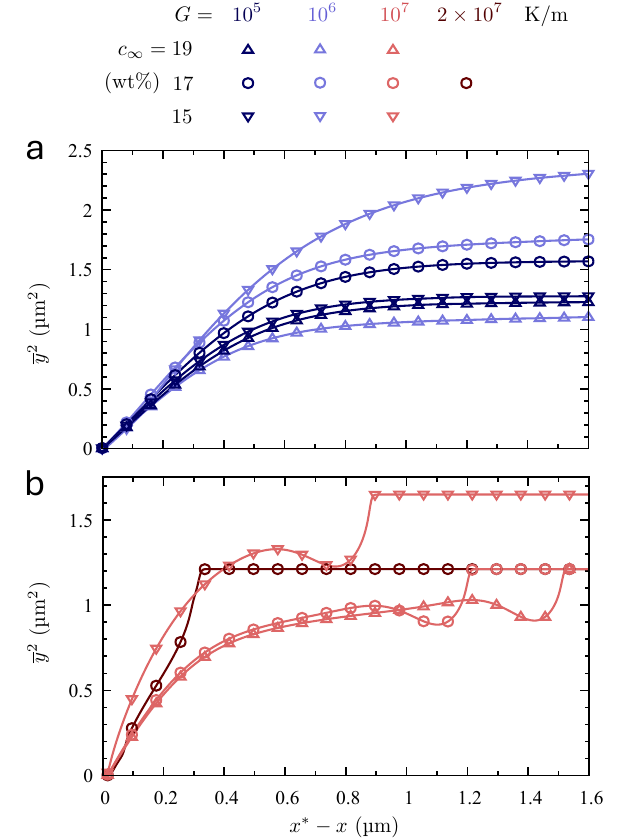}
\caption{
Average square width $\overline y^2$ of cellular/dendritic features, calculated as $\overline{y}(x) = \int\{(1+\phi(x,y))/2\}{\rm d}y/N$, with $N$ the number of cells or dendrites in the simulation, as a function of the distance to the tip, $x^*-x$, for all non-planar PF cases at $V=0.012$~m/s and $c_\infty=17$\,wt\%: (a) dendrites at $G\leq10^6$~K/m, (b) cells at $G\geq10^7$~K/m.
}
\label{fig:dend_or_cell}
\end{figure}

The main effect of $G$ appears on the microstructure morphologies.
As expected, a lower $G$ promotes the growth of dendritic microstructure, while high-$G$ microstructures appear more cellular.
This is can be associated to the fact that a given growth velocity $V$ gets closer to the low-$V$ constitutional undercooling threshold $V_c$, which increases with $G$.
This morphological transition, particularly apparent at low $V=0.012$~m/s, can be ascertained objectively by plotting the cell/dendrite cross-section area (or, in 2D, width squared, here noted $\overline y^2$) as a function of the distance to the tip, as in Fig.~\ref{fig:dend_or_cell}.
The parabolic shape of a dendrite is expected to result in a linear dependence of $\overline y^2$ as a function of the distance to the tip $x^*-x$ in the vicinity of the tip ($x^*-x\approx0$).
Figure~\ref{fig:dend_or_cell}a includes patterns identified as dendritic, for $G\leq10^6$~K/m, showing a linear slope in the vicinity of the tip, while Fig.~\ref{fig:dend_or_cell}b shows nonlinear profile associated to cells at $G\geq10^7$~K/m.

At low $G$ and low $V$, e.g. panels (a$_1$) in Figs.~\ref{fig:C15grid}-\ref{fig:C19grid}, microstructures exhibit dendritic features (e.g. a parabolic tip).
Still at low $G$, as $V$ increases, first the density of doublons  increases.
A doublon  is composed of two asymmetric fingers, with a mirror symmetry between them, growing cooperatively with a narrow liquid channel along the symmetry axis.
They constitue a known steady-state solution of the underlying free-boundary problem, characteristic of high undercooling and low interface anisotropy \cite{brener1992kinetic, ihle1994fractal, amar1995parity, akamatsu1995symmetry, kupferman1995coexistence} --- here possibly promoted by the fact that simulations are two-dimensional.
As $V$ increases further, still at low $G$, the microstructure adopts a peculiar ``wavy'' morphology (see, e.g., Figs~\ref{fig:C15grid}a$_5$ and \ref{fig:C19grid}a$_5$) just before the transition to a planar growth.
Some of these patterns (further below $V_a$) exhibit clearly marked spacings, which, while still measurable, may be notably heterogeneous across the array (Fig.~\ref{fig:C19grid}a$_5$).
When $V\to V_a$, the pattern becomes increasingly unstable and the lateral movement of the interface ``hills'' and ``valleys'' leaves behind tortuous segregation patterns (see Fig.~\ref{fig:C15grid}a$_5$, as well as panels a$_1$, a$_4$, b$_1$, b$_4$, c$_1$, and c$_4$ in Fig.~\ref{fig:cellplanarspace}).
This latter type of microstructure was identified as ``intermediate'' in the classification algorithm, due to the impossibility to define a clear steady primary spacing.
Additional PF simulations performed for extended periods of time have confirmed that these intermediate structures do not eventually stabilize, reinforcing their distinction from stable steady-state cells or dendrites.

At higher $G$, e.g. for $G=10^7$\,K/m, microstructures deviate from dendrites and exhibit cellular features, with a rounded tip shape and shallower intercellular liquid channels.
In this cellular regime, the transition from cellular to planar as $V$ increases occurs with a reduction of both primary spacing and liquid channel depth (i.e. cells become increasingly shallower), but apparently no unstable ``intermediate'' regime comparable to that observed for lower $G$.

The use of the frozen temperature approximation, Eq.\,\eqref{FTA}, may introduce limitations that should be considered. 
Indeed, at high solidification velocity, the rate of release of latent heat, and its subsequent diffusion in the liquid, may become important.
This is known to strongly affect the region of occurrence (e.g. onset composition) of oscillatory banding instability, and the resulting microstructural length scales (i.e. band spacing) --- as theorized via linear stability analysis \cite{karma1993interface} and recently confirmed by PF simulations \cite{ji_quantitative_2024}. 
As we selected an alloy that does not exhibit any banding instability, we expect a weaker effect of the latent heat release, and that the observed trends remain valid if one considers the imposed $G$ as an effective temperature gradient in the vicinity of the interface. 
Still, a deeper quantitative study of the effect of latent heat release on the present results remain needed in the future. 

\subsubsection{Effect of the diffuse interface width $W$}

Beyond the effect of physical parameters $(G,V,c_\infty)$, our study also explores the effect of the diffuse interface width $W=SW_0$ in the selected PF model, by comparing Fig.~\ref{fig:C17grid} (performed with $S=5$) and Fig.~\ref{fig:S8grid} (performed with $S=8$), both with $c_\infty=17$\,wt\%\,Cr.

The most striking difference appears close to the absolute stability threshold at $V=0.3$~m/s, where the transition to planar growth at high $G$, visible at $S=5$ (Fig.~\ref{fig:C17grid}c$_5$) is absent for $S=8$ (Fig.~\ref{fig:S8grid}c$_5$).
This suggests that using a wider diffuse interface in the PF model promotes cellular growth at the expense of planar growth in the vicinity of the transition. 

In terms of PDAS, the two cases also exhibit differences. The $S=8$ simulations generally slightly overestimate the dendritic spacing compared to $S=5$. 
This discrepancy may be attributed to an increase in grid-related anisotropy associated with a larger interface width.

Moreover, the presence of ``wavy'' interdendritic liquid channels seems to be promoted by the use of a wider interface -- see, e.g., Fig.~\ref{fig:S8grid}b$_2$, b$_3$ and a$_2$. 
For $S=5$, such wavy interdendritic channels are only observed for $V=0.06$~m/s and $G=10^6$~K/m (Fig.~\ref{fig:C17grid}b$_3$), indicating that this morphological feature is a numerical artifact associated with increased interface thickness. Indeed, an additional simulation for these conditions with $S=3$ confirmed that these wavy intercellular channels completely straighten with a narrower diffuse interface. This higher-resolution simulation required almost 7 days (wall time) on 4 GPUs, which is significantly more computationally intensive than the $S=5$ case (5 days on 2 GPUs). Keeping in mind that simulations at lower $G$ values would impose an even greater computational cost, the choice of $S=5$ results in a good balance between accuracy and computational cost. 

In summary, even though microstructures at $S=5$ and $S=8$ are very close, highlighting the robustness of the model using an upscaled interface with $S>1$, these observations also highlight the importance of the interface width parameter in PF simulations, as it directly impacts the morphology, stability, and characteristic spacing of dendritic growth.

\subsection{Bayesian active learning}

A complete -- e.g. full factorial -- mapping of $(G,V,c_\infty)$ conditions as presented above provides valuable information on microstructure selection in rapid solidification. 
However, it is costly, inefficient, and its scope is drastically limited compared to what would be required in an actual technological application like AM of multicomponent alloys -- in particular, it is restricted to binary alloys under simplified thermal conditions. 
Therefore, efficient exploration schemes for high-dimensional parameter spaces are required, such as the Bayesian-guided active learning scheme introduced in Section~\ref{method:bayesian}, and discussed below.

\subsubsection{Final selection map} \label{disc:finalmap}

Figure~\ref{fig:cellplanarspace} presents the final results of the Bayesian-guided PF exploration, including results of all simulations conducted in this study, of which an exhaustive list is provided in the attached Supplementary Material. 
The results confirm the presence of a transition from dendritic/cellular growth (purple) at moderate growth velocity $V$ to planar growth (green) at high $V$. 

Lower alloy concentrations ($c_\infty$) lead to a lower absolute stability threshold, which is consistent with the initial batch of PF simulations (Figs~\ref{fig:C15grid}-\ref{fig:C19grid}), as well as with using Eq.~\eqref{eq:absolutestability} or the value of $\max\{T_S(V)\}$ from Eq.~\eqref{tsolidus} as the morphological transition threshold (see Section~\ref{disc:conc}).

Moreover, while these simple theories do not include an explicit dependence of the transition velocity upon the temperature gradient $G$, our results show that a higher $G$ also tends to promote planar solidification. 
As discussed in Section~\ref{disc:G}, this is already apparent from the initial batch of PF simulations (Figs~\ref{fig:C15grid}-\ref{fig:C19grid}), where the velocity range for the transition from patterned to planar growth changes from $0.3<V_a {\rm (m/s)}<0.6$ at $G=10^5$\,K/m to $0.12<V_a {\rm (m/s)}<0.3$ at $G=10^7$\,K/m, for all values of $c_\infty$ (with $S=5$).

Interestingly, $G$ also appears to have a strong effect (in combination with $c_\infty$) on the width of the velocity range for this transition.
Indeed, low $G$ and high $c_\infty$ seem to promote the formation of structures that could not be unambiguously classified as either cellular/dendritic or planar, here referred to as ``intermediate'', represented by the white region in Fig.~\ref{fig:cellplanarspace} and illustrated in panels a$_1$, a$_4$, b$_1$, b$_4$, c$_1$, and c$_4$ therein.
As discussed in Section~\ref{disc:G}, such intermediate structure emerges just below $V_a$ when the moderate-$V$ morphology is dendritic (but not cellular).
Such unstable microstructures substitute the oscillatory banding regime observed in other alloys, e.g. Al \cite{carrard1992banded, karma1993interface, kurz1996banded, mckeown2016time} or Mg alloys \cite{tourret2023morphological}.
The absence of such a banding regime in the present Fe-Cr alloy can be related to its high level of kinetic undercooling relative to the effect of solute trapping, resulting in a faint peak of the $T_S(V)$ curve (Fig.~\ref{fig:spacings}) -- even more attenuated considering $k_V$ and $m_V$ corresponding to the PF model (pink dotted lines) -- and hence to the resulting absence of a significant unstable $V$ range with $dT_S/dV > 0$.

\subsubsection{Active learning through iterations}

\begin{figure}[b!]
\centering
\includegraphics[width=\columnwidth]{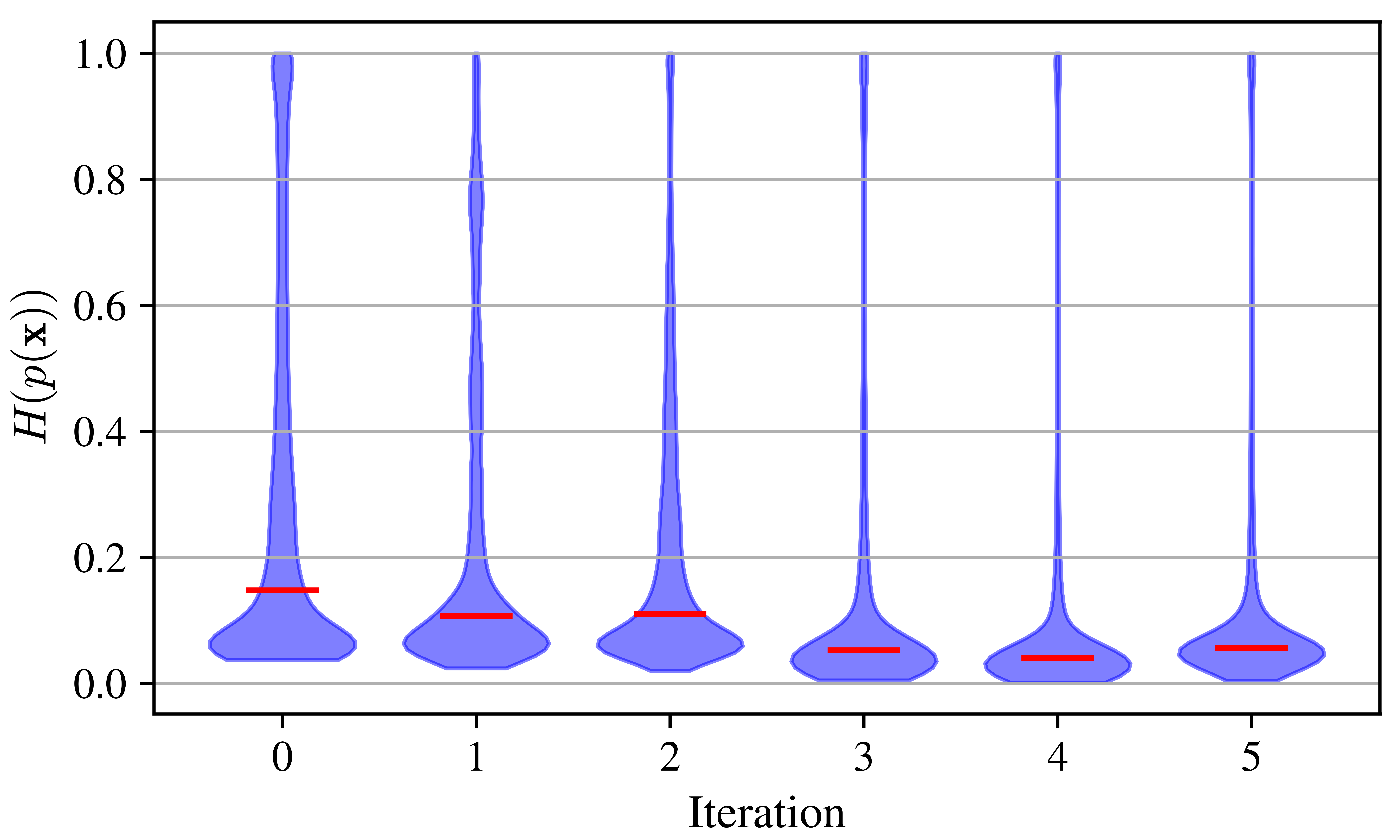}
\caption{
Violin plots showing the distribution and median of Shannon entropy values within the $(G,V,c_\infty)$ domain as a function of iteration. Higher values of entropy correspond to higher uncertainty while lower values correspond to lower uncertainty. 
The slight increase in the Shannon entropy between 4th and 5th iteration is due to the length scales being manually adjusted in the final probability mapping (see Section~\ref{PriorMeanFunc}).
}
\label{fig:H_red}
\end{figure}

Figure \ref{fig:iterations_grid} illustrates the evolution of the Gaussian process model over successive learning iterations, guided by monitoring the Shannon entropy $H(p(\mathbf{x}))$ across the design space $\mathcal{X}$.
We begin with a prior based on $V_{\max\{T_S\}}$ for all $G$ (top row), and a 0$^{\rm th}$ seed round based on 54 initial points from full factorial PF mapping, which generates an approximate map of the class boundaries. While this first mapping captures general trends — such as a tendency for planar morphologies at high $V$ and high $G$ — it lacks precision, especially near regime transitions, and contains large regions of uncertainty. This uncertainity corresponds to higher information entropy, as shown in the early iterations in Figure \ref{fig:H_red}. 

As new data is incorporated over successive iterations, we observe significant refinement of the decision boundaries and reduction in information entropy (Figure \ref{fig:H_red}). After the first iteration, the inclusion of the initial data significantly sharpens the boundary between planar and cellular microstructures. However, uncertainty remains substantial in areas where the model has not yet observed representative intermediate structures — particularly in high-concentration, low-gradient regions and near $V = 0.3$~m/s.

As iterations progress, especially after iteration 2, the model encounters intermediate microstructures in this uncertain region, leading to increased predictive entropy. The acquisition function, guided by Bayesian uncertainty, effectively directs sampling toward this area in subsequent iterations. By the final iteration, this adaptive sampling strategy results in a much more refined prediction of class boundaries, as shown in Figure \ref{fig:H_red}. This illustrates the effectiveness of Bayesian updating in identifying and resolving, particularly in critical transition zones.

\begin{figure*}[t!]
\centering
\includegraphics[width=.8\textwidth]{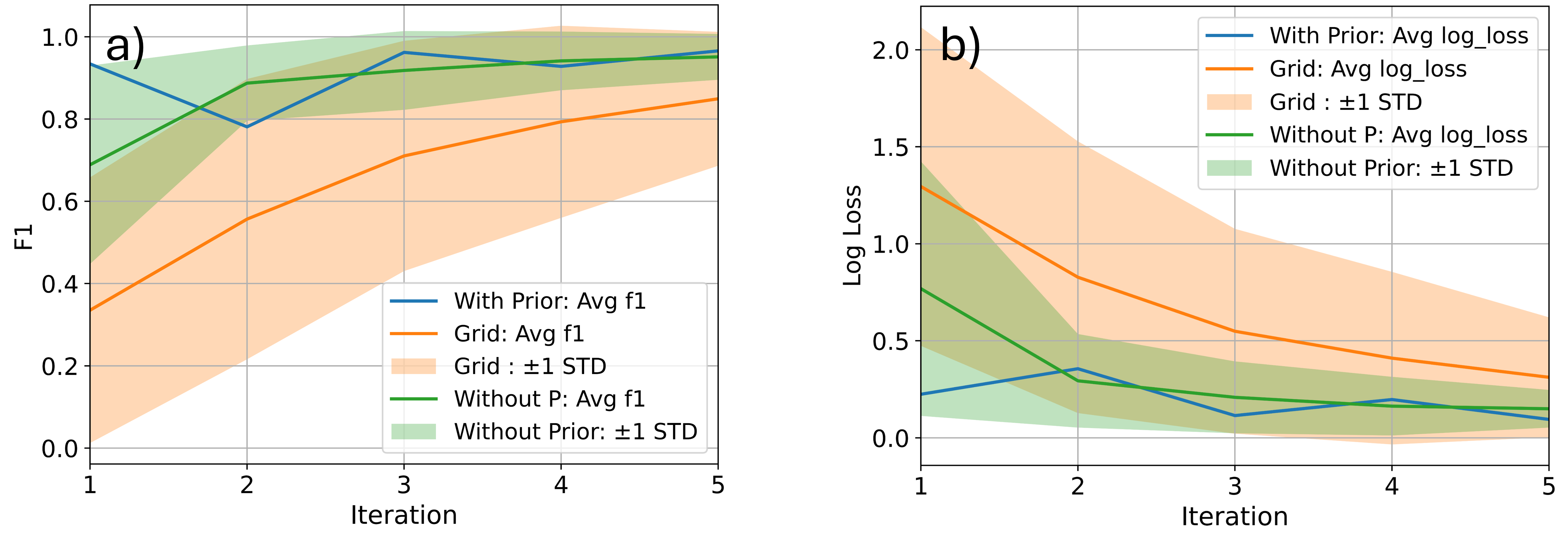}
\caption{
Results of retrospective performance analysis of different sampling and learning strategies.
a) The F1 score distributions as a function of iteration captures the ability of the model to assign correct solidification morphologies within the $(G,V,c_\infty)$ space. b) The Log-Loss distributions captures the calibration of the predicted class probabilities, i.e., it penalizes overconfident or poorly calibrated predictions even when the final class label is correct. To ensure statistical robustness, each acquisition method was independently repeated over 200 runs -- shaded areas illustrate the region with plus/minus one standard deviation. 
}
\label{fig:in_silico_bench}
\end{figure*}

To quantitatively demonstrate the Bayesian approach efficency, we performed an \emph{in silico} retrospective analysis comparing the effectiveness of different exploration strategies. Specifically, we treat the GP classifier trained after 5 iterations (Figure \ref{fig:iterations_grid}) as the ground truth. This assumes that the classifier, trained using 104 PF simulations, serves as a high-fidelity emulator of the solidification morphology and can reasonably replace direct PF simulations for benchmarking purposes. Using this ground truth GP classifier, we evaluated 3 acquisition methods: (1) Bayesian active learning with the CGM serving as an informative prior, (2) Bayesian active learning with an uninformative prior, and (3) grid-based sampling, in which points are randomly selected from a discretized grid spanning the process space. The grid was defined over the $(G,V,c_\infty)$ domain, with both velocity ($V$) and gradient ($G$) logarithmically sampled. These will be referred to as campaigns 1-3.

For all campaigns, we employed the Shannon Entropy + KMedoids acquisition function. To initialize Campaigns 2 and 3, we began with 5 randomly selected points within the $(G,V,c_\infty)$ domain. To initialize Campaign 1 we used the Shannon Entropy + KMedoids acquisition function on the prior, i.e., we initialized the campaign with data according to where the CGM predicts the boundary to be. In doing so, this makes the sequence of recommended experiments in Campaign 1 deterministic. A batch size of 5 was chosen to accentuate the performance differences between methods; larger batches tend to drive all strategies rapidly toward convergence. To further highlight differences between the informative and uninformative priors, we increased the influence of the CGM prior: if the CGM predicted a point to be planar, we assigned a prior belief of $p_p = 90\%$ for it being planar and $p_p = 10\%$ for being dendritic/cellular, and {\it vice versa} when the CGM predicted dendritic/cellular.

At each iteration, the data acquired through each method was used to train a surrogate GP classifier, which was then evaluated against the ground truth classifier. Performance was assessed using both the F1 score and log-loss. The F1 score was chosen because it provides a single, balanced measure of precision and recall \cite{hardcastle2025physics}. The log‑loss was selected because it directly evaluates the quality and calibration of the predicted class probabilities, penalizing overconfident or poorly calibrated predictions even when the final class label is correct \cite{hardcastle2025physics}. To ensure statistical robustness, each acquisition method was independently repeated over 200 runs, resulting in 200 error metrics per iteration. 

Results of this \emph{in silico} retrospective analysis are summarized in Figure~\ref{fig:in_silico_bench}.
It clearly shows that Bayesian methods perform better in terms of both F1 score (Fig.~\ref{fig:in_silico_bench}a) and log-loss (Fig.~\ref{fig:in_silico_bench}b) metrics.
The method using a more informative prior shows an advantage at first iteration, however it vanishes quickly in successive iterations.
Beyond average performance, the wide standard deviations observed in Campaign 3 indicate that, as expected, a non-guided sampling method is highly variable across runs. In contrast, Campaign 2 (uninformative‑prior Bayesian learning) exhibits much narrower error bands, reflecting more consistent outcomes regardless of initial seedling data. 
Campaign 1 (informative CGM prior) shows zero standard deviation in F1 score, because its initial seedling data is drawn deterministically from the CGM, hence every run follows an identical query sequence. Although the mean performance of Campaign 1 closely parallels that of Campaign 2 at each iteration, the informative prior offers a clear advantage in that it does not significantly degrade predictive accuracy (even in the worst‑case at iteration 2 in Figure \ref{fig:in_silico_bench}a,b) yet results in steady, repeatable improvement in model quality. When average performance is nearly the same, we prefer the surrogate model that results in more consistent performance during active learning campaigns \cite{hardcastle2025physics}. 

Overall, the Bayesian framework — combining a Gaussian Process model with an uncertainty-driven acquisition function — proved to be an efficient strategy for exploring the $(G,V,c_\infty)$ parameter space. 
By leveraging probabilistic predictions and targeting high-entropy regions, the model focused data acquisition in areas that were both informative and uncertain. 
The inclusion of a clustering-based selection mechanism further ensured spatial diversity among sampled points, enabling broad coverage of the input space without excessive redundancy. 
As a result, the model was able to learn the most critical features of the dendritic/cellular-planar morphological transition with just 104 simulations, with all of the 50 Bayesian-guided PF simulations focused on regions of highest interest --- as illustrated, for instance, by the high density of data points close to the cellular/planar boundary in Fig~\ref{fig:gv}. 
This highlights the power of Bayesian active learning to effectively allocate computational resources and accelerate discovery in high-dimensional physical design spaces.

\section{Conclusions}

Here, we used phase-field simulations, guided by Bayesian active learning, to investigate microstructure selection in rapid solidification of a binary Fe-Cr alloy, a surrogate binary alloy for a 316L stainless steel, under conditions relevant to additive manufacturing. 
Our results highlight the complex interplay between solidification velocity ($V$), thermal gradient ($G$), and alloy composition ($c_\infty$) in determining the resulting microstructural morphologies. 
The simulations capture the transition from dendritic/cellular growth at moderate velocities to planar growth at higher velocities. This aligns with fundamental solidification theory, but the PF model captures nuanced deviations from analytical predictions, and provides deeper insight into the morphologies of resulting microstructures. 

The range of $V$ explored here covers the transition from fully cellular/dendritic to planar growth.
The range of $G$ cover the transition from dendritic (low $G$) to cellular (high $G$) to fully planar across the entire $V$ range (highest $G$).
The transition from dendrites to cells with increasing $G$ appears related to the proximity of the low-$V$ constitutional undercooling (or Mullins-Sekerka) instability velocity with the high-$V$ absolute stability threshold.

Our results seem to discard the presence of a clear, steady, and stable intermediate cellular regime between dendritic and planar growth at high $V$ and low $G$. 
Instead, the microstructures exhibit an unstable ever-evolving ``wavy'' pattern, labelled here as ``intermediate''.
These intermediate structures do not stabilize over extended simulation times, suggesting a dynamic competition between different growth modes, in contrast to the smooth transition observed in the cellular regime at high $G$. 
The presence of this regime is linked to the absence of banding in the Fe-Cr system --- due to the relatively weak effect of solute trapping compared to that of kinetic undercooling in this alloy, leading to the vanishing of the $V$ range with ${\rm d}T/{\rm d}V>0$ (Fig.~\ref{fig:spacings}). 

Comparing PF results with classical theories, we found that the calculated value of $G$ above which planar growth is stable across the entire $V$ range is precisely matched between PF and KGT models.
The dependence upon $G$ of the absolute stability threshold velocity is also consistent between PF and KGT models.
However, its value at low $G$ is better predicted by using the maximum of the $T_S(V)$ curve, here denoted $V_{\max\{T_S\}}$, rather than $V_a$ from the classical formula, Eq.~\eqref{eq:absolutestability}.
A simple shift of the KGT-predicted absolutely stability limit by a factor $V_{\max\{T_S\}}/V_a$ provides a simple and accurate estimate of the actual absolute stability limit assessed by Bayesian-guided PF simulations (Fig.~\ref{fig:gv}).
We also found that the KGT model and theoretical solidus temperature $T_S(V)$ provide a reasonable estimate of the solidification front temperatures in the dendritic/cellular and in the planar regime, respectively.
At a given $G$, in the dendritic regime, PF-predicted primary arm spacing exhibit an increase as $V\to V_a$, akin to the evolution of the tip radius calculated via the KGT model.
An increase of $G$ leads to a transition from dendritic to cellular, and a greater discrepancy with the KGT model (based on the assumption of dendritic growth).
Further analysis remains needed in order to identify the effect of latent heat release (i.e. the validity of the frozen temperature approximation), as well as the effect of different material parameters, such as kinetic coefficient and interfacial anisotropies, on the present results.
Preliminary results from an ongoing study suggest that the most important effect of these parameters relate to the occurrence (or inhibition) of the unstable banding regime and its oscillation characteristics (e.g. temperature and velocity ranges covered by the oscillations).

The methodology presented here, combining phase-field simulations and Bayesian active learning, resulted in an efficient mapping of solidification regimes and revealed complex microstructural behavior, including unstable intermediate growth patterns. 
Provided an appropriate quantitative PF model, our Bayesian-guided exploration strategy can be applied to other alloy systems where a precise control of microstructural development is critical for achieving desired material properties. Further investigation into the intermediate growth regime is warranted to fully understand its generality, its conditions of occurrence, and its potential influence on material properties and performance. Further ongoing investigation also includes the identification of key material properties that promote the occurrence and characteristics of banded microstructures in the dendritic-to-planar velocity regime.

\section*{Acknowledgments}
JM, RT, and DT acknowledge financial support from the Spanish Ministry of Science via a Mar\'ia de Maeztu seal of excellence [CEX2018-000800-M] (JM), international collaboration project MiMMoSA [PCI2021-122023-2B] (RT) and a Ram\'on y Cajal grant [RYC2019-028233-I] (DT). 
RA and JM acknowledge support from AFRL through a sub-contract with ARCTOS, TOPS VI (165852-19F5830-19-02-C1). 
RA and BV also acknowledge the Data-Enabled Discovery and Design of Energy Materials (D$^3$EM) program funded by NSF through Grant No.~DGE-1545403. 
RA also acknowledges partial support from DARPA under Grant No.~HR0011-25-2-0009. 
JM was also supported by NSF through Grant No.~2328395. 
Simulations were carried out at the Texas A\&M High Performance Research Computing (HPRC) Facility.

\section*{CRediT author statement}

{\bf Jos\'e Mancias}: Conceptualization; Methodology; Software; Validation; Formal analysis; Investigation; Data Curation;
Writing -- Original Draft; Visualization.
{\bf Brent Vela}: Conceptualization; Methodology; Writing -- Original Draft. 
{\bf Juan Flórez-Coronel}: Conceptualization; Methodology; Software; Visualization.
{\bf Rouhollah Tavakoli}: Methodology; Software; Formal analysis; Writing - Review \& Editing.
{\bf Douglas Allaire}: Writing -- Review \& Editing; Funding acquisition.
{\bf Raymundo Arr\'oyave}: Writing -- Review \& Editing; Supervision; Funding Acquisition
{\bf Damien Tourret}: Conceptualization; Methodology; Formal analysis; Writing -- Review \& Editing; Visualization; Supervision; Funding acquisition.

\section*{Data Availability}
Data will be made available on request. 

\appendix
\setcounter{figure}{0}

\section{Phase Field Convergence Analysis} \label{appdx:convergence}

\begin{figure*}[ht!]
  \centering
  \includegraphics[width=\textwidth]{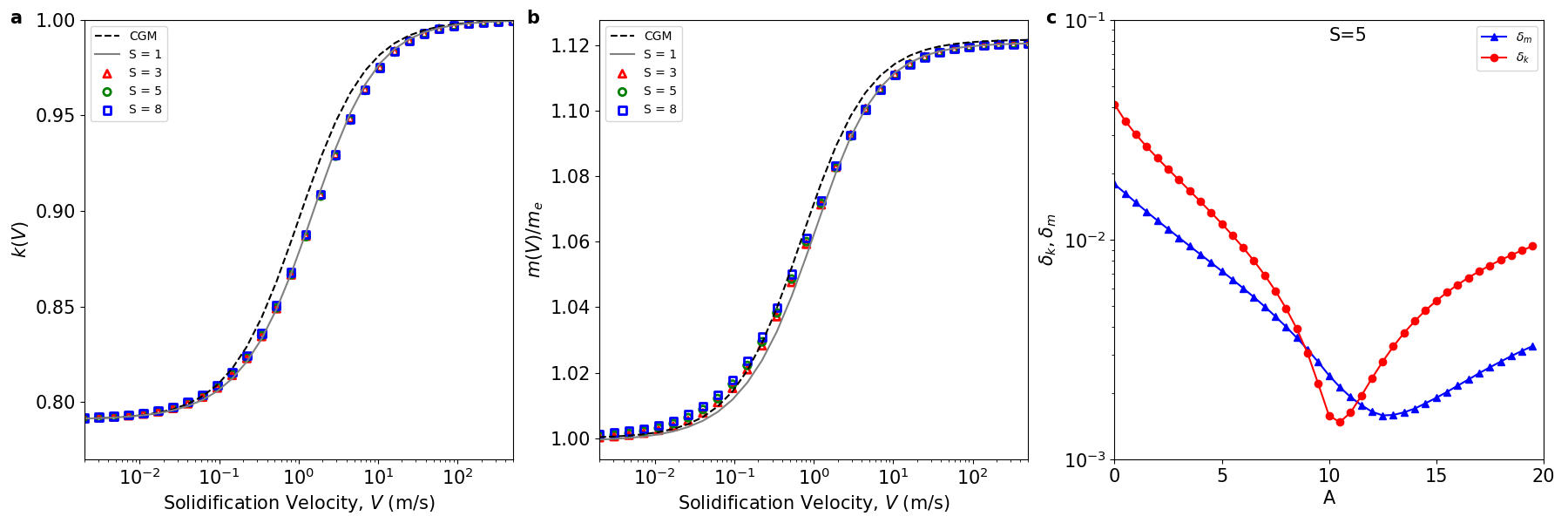}
  \caption{Results from preliminary PF calibration and verification steps, using a 1D formulation considering a steady-state profile $\phi_0(x)$. Plots of (a) partition coefficient and (b) liquidus slope as a function of solidification velocity for (S,A) = (1,1) in the solid grey line and (S,A) = (3,6), (5,11), (8,18) are in colored symbols and compared to the CGM predictions from equations \ref{eq_kV} and \ref{eq_mVme} in the dashed black line. (c) Shows the calculated deviations $\delta_k$ and $\delta_m$ integrated over the entire $V$ range between PF-predicted $k_V$ and $m_V$ and the reference solution at $S=1$ and $A=1$, here $S=5$ is shown. 
  }\label{fig:APPX_S_A}
\end{figure*}

Using the PF model and parameters outlined in Section~\ref{method:phasefield}, we performed a preliminary convergence analysis with respect to $S=W/W_0$ in order to identify appropriate values of the interface enhanced diffusion parameter $A$, as proposed in \cite{ji_microstructural_2023,ji_quantitative_2024} and also performed in \cite{tourret_emergence_2024}.
To select $A$ for a given $S$, we compute the $k_V$ and $m_V$ curves for the reference case at $S=A=1$ over a wide range of $V$. 
Assuming that the phase field $\phi$ closely follows its theoretical stationary profile $\phi_0 = -\tanh[x/(\sqrt{2}W)]$, the corresponding concentration profile $c(x)$ can be determined by numerical integration of 
\begin{equation} \label{appx:dcdx}
    \frac{dc}{dx} = (c_\infty - c) \frac{V}{D_L q(\phi)} + bc \frac{dg(\phi)}{dx}
\end{equation}
using the boundary conditions $c(\pm \infty) = c_\infty$. 

From the calculated $c(x)$ profile at a given $V$, the partition coefficient $k_V$ is determined by the ratio of the concentrations at the solid ($c_s$) and liquid ($c_l$) sides of the interface, respectively approximated as the nominal concentration $c_\infty$ and the maximum of $c(x)$. The slope of the liquidus line is then obtained by numerical integration of
\begin{equation} \label{appx:mvme}
    \frac{m_V}{m_e} = \frac{b}{(1-k_e)c_l} \int^{+\infty}_{-\infty} g'(\phi)c \frac{d\phi}{dx}dx
\end{equation}

After computing $k_V$ and $m_V$ across the whole velocity range for a given set of $(S,A)$ values, their deviation from the reference case at $S=A=1$ is quantified as $\delta_k$ and $\delta_m$. These are respectively calculated as the discrete averages of $\sqrt{(k_S-k_1)^2}/k_1$ and $\sqrt{(m_S - m_1)^2}/m_1$ over the calculated velocity range, with $k_1$ and $m_1$ the reference values of $k_V$ and $m_V$ at $S=A=1$ and $k_S$ and $m_S$ the values calculated for $S \neq 1$ \cite{ji_quantitative_2024}. 

Numerical integration of equations \eqref{appx:dcdx} and \eqref{appx:mvme} was performed up to machine precision using a custom python script for 30 different velocities equally spaced in log-space between $10^{-2.7}$ m/s and $10^{2.7}$ m/s. The optimal value of $A$ is then selected by minimizing the averages of $\delta_{k}(A)$ and $\delta_{m}(A)$ as illustrated in Figure~\ref{fig:APPX_S_A}. 

In addition to this simple analysis, which considers an approximate stationary profile $\phi_0 = -\tanh[x/(\sqrt{2}W)]$, we also performed the calculations of 1D steady state solutions of the coupled PF equations in a moving frame, as described in Appendix C of \cite{ji_quantitative_2024}.
The resulting ($k_V$, $m_V$), which fall very close to the approximate results in Fig.~\ref{fig:APPX_S_A}, were used to calculate the PF-consistent solidus line, $T_S(V)$, in Fig.~\ref{fig:spacings}.

\bibliography{citations}

\end{document}